\documentclass[12pt]{article}
\usepackage[pdftex]{graphicx,color}
\usepackage{amsmath,amssymb,booktabs,enumerate}
\usepackage{subfigure, url}
\usepackage{comment}
\usepackage{ulem}
\def\a{\alpha}

\def\e{\epsilon}

\def\g{\gamma}

\def\l{\lambda}
\def\m{\mu}
\def\n{\nu}

\def\q{\partial}

\def\s{\sigma}
\def\t{\tau}

\def\D{\Delta}
\def\G{\Gamma}

\def\beq{\begin{eqnarray}}
\def\eeq{\end{eqnarray}}
\newcommand{\gsim}{ \gtrsim}
\newcommand{\lsim}{ \lesssim}

\newcommand{\GEV}{ {\rm GeV} }
\newcommand{\TEV}{ {\rm TeV} }

\def\sla#1{\rlap/#1}
\begin{document}
\baselineskip 0.7cm
\begin{titlepage}

\begin{flushright}
KEK-TH-1773
\end{flushright}

\vskip 1.35cm
\begin{center}
{\large \bf 
Discrimination of dark matter models in future experiments
}

\vskip 1.2cm
Tomohiro Abe$^1$, Ryuichiro Kitano$^{1,2}$ and Ryosuke Sato$^1$
\vskip 0.4cm

{\it
$^1$Institute of Particle and Nuclear Studies,\\
High Energy Accelerator Research Organization (KEK)\\
Tsukuba 305-0801, Japan\\
$^2$The Graduate University for Advanced Studies (Sokendai)\\
Tsukuba 305-0801, Japan\\
}

\vskip 1.5cm

\abstract{
Phenomenological aspects of simple dark matter models are studied.
We discuss ways to discriminate the dark matter models in future experiments.
We find that the measurements of the branching fraction of the Higgs boson
 into two photons and the electric dipole moment of the
 electron as well as the direct detection experiments are quite useful
 in discriminating particle models of dark matter. We also discuss the
 prospects of finding new particles in dark sector at the LHC/ILC.
}
\end{center}
\end{titlepage}
\setcounter{page}{2}

\tableofcontents

\section{Introduction}

Dark matter (DM) was first proposed by Oort \cite{Oort} and Zwicky
\cite{Zwicky:1933gu} 
to explain the motion of stars in our galaxy or galaxies in clusters.
Eighty years have passed since then, and various evidences
(e.g., galaxy rotation curves, gravitational lensing, precision
measurement of cosmic microwave background and so on) support the
existence of the dark matter. 
However, we still do not know what the dark matter is.
Various candidates have been proposed. The most promising one is weakly
interacting massive particle (WIMP) \cite{Steigman:1984ac}. 
Since there is no candidate of such a particle in the standard model
(SM), this scenario requires an extension of the SM.

The recent discovery of the Higgs boson and the measurements of its
properties strongly support the origin of the Higgs boson as a component of
the $SU(2)_L$ doublet Higgs field.
Precise measurements of its properties to
understand the nature of the Higgs field are one of the most important
tasks in particle physics. It is quite possible that the DM particle
couples to the SM through the Higgs field so that the DM abundance is
explained as thermal relic {\it \`a la} the WIMP scenario. In that
case, the Higgs boson as well as other SM particles carries information
on the DM. In literature, such kind of various particle models have been
proposed and their phenomenology have been studied, for
example~\cite{Silveira:1985rk, McDonald:1993ex,
Burgess:2000yq,Cline:2013gha, 
Barbieri:2006dq, Arhrib:2013ela, Krawczyk:2013wya, Belanger:2013xza,
Araki:2011hm, Cirelli:2007xd, 
Cohen:2011ec, Cheung:2013dua,
Dedes:2014hga}.

In this paper, we survey simple extensions of the SM to account for dark
matter of the Universe by the WIMP scenario, and summarize the current
situations and future prospects to observe signatures of each model. 
We list six renormalizable models to realize the WIMP scenario as examples,
and compare the model predictions to see if we can distinguish models by
various measurements. 
We review and summarize the status of each model thoroughly, and also
show new results such as the predictions to the electron electric dipole moment (EDM)
 in the fermionic dark matter models. We examine which observables are important in each models, and discuss the differences.

%
%
This paper is organized as follows.
In section \ref{sec:model}, we introduce dark matter models which we will
discuss in this paper. 
In section \ref{sec:pheno}, we briefly review phenomenology in each
models, especially focusing on the spin-independent cross section, Higgs
invisible decay, Higgs diphoton signal, and electron EDM. 
We impose in each model that the energy density of the dark matter abundances 
$\Omega_{\rm DM} h^2 = 0.1196 \pm 0.0031$ which is reported by the
Planck collaboration \cite{Ade:2013zuv} is explained.
In section \ref{sec:discriminate},
we discuss how to discriminate models in future experiments
Section~\ref{sec:summary} is devoted for conclusion and discussions.

\section{Dark Matter Models}
\label{sec:model}
In this section, we list dark matter models which we discuss in this paper.
We add new field(s) to the SM and introduce $Z_2$ parity which
guarantees the stability of the dark matter.
Under this $Z_2$ parity, all of the SM fields are even and the new fields
are odd. We take minimal renormalizable Lagrangian which includes a
candidate of dark matter. We summarize the models in Tab. \ref{tab:modellist}.

\begin{table}
\begin{center}
\caption{List of dark matter models.
Numbers in a parenthesis attached to the field represent $SU(2)_L \times U(1)_Y$ representation.
For the Lagrangian of each model, see each section.}
 \label{tab:modellist}
\begin{tabular}{|c|c|c|}
\hline
Model & $Z_2$ odd field(s) & Parameters\\
\hline\hline
S1 (Sec. \ref{subsec:s1}) & $s(1_0)$ & $m_s$, $\l_{sH}$\\
\hline
S2 (Sec. \ref{subsec:s2}) & $H_2(2_{1/2})$ & $m_{A^0}$, $m_{S^0}$, $m_{H^+}$, $\l_A$, $\l_2$\\
\hline
F12 (Sec. \ref{subsec:f12}) & $\psi_S (1_0), \psi_D(2_{-1/2}), \chi_D(2_{1/2})$ & $m_S$, $m_D$, $\l$, $\l'$, $\theta$\\
\hline
F23 (Sec. \ref{subsec:f23}) & $\psi_D (2_{-1/2}), \chi_D(2_{1/2}), \psi_T(3_0)$ & $m_D$, $m_T$, $\l$, $\l'$, $\theta$ \\
\hline
\end{tabular}
\end{center}
\end{table}

\subsection{Singlet scalar dark matter (Model S1)} \label{subsec:s1}
In this model, an additional $SU(2)_L$ singlet real scalar $s$ with hypercharge $Y=0$ is introduced \cite{Silveira:1985rk, McDonald:1993ex, Burgess:2000yq}.
Mass and interaction terms for $s$ are given by,
\begin{align}
{\cal L}_{\rm S1} &=  - \frac{m_1^2}{2} s^2 - \frac{\l_{sH}}{2} s^2
 |H|^2 - \frac{\lambda_s}{4!} s^4.
\end{align}
The self-interaction term, $s^4$, does not affect following discussion.
The mass eigenvalue of $s$ is given by $m_s^2 = m_1^2 + \l_{sH}v^2/2$,
where $v \simeq 246~\GEV$ is the VEV of the Higgs field.

\subsection{Doublet scalar dark matter (Model S2)} \label{subsec:s2}
In this model, an additional $SU(2)_L$ doublet scalar $H_2$ with
hypercharge $Y=1/2$ is introduced \cite{PHRVA.D18.2574, Barbieri:2006dq}.
Mass and interaction terms for $H_2$ are given by,
\begin{align}
{\cal L}_{\rm S2} &= - m^2_2 |H_2|^2 - \l_1|H|^4 - \l_2|H_2|^4 - \l_3|H|^2|H_2|^2 - \l_4|H^\dagger H_2|^2 - \frac{\l_5}{2}\bigl[ (H_2^\dagger H)^2 + h.c. \bigr].
\end{align}
In general, $\l_5$ is a complex parameter, however, its phase can be taken away by a redefinition of $H_2$.
In the following of this paper, we take $\l_5$ as real and positive. 
$H_2$ is decomposed as,
\begin{align}
H_2 =
\left(
\begin{array}{c}
H^+ \\
(S^0+iA^0)/\sqrt{2}
\end{array}
\right),
\end{align}
where $H^+$ is a charged scalar field and $S^0$ and $A^0$ are neutral
real scalar fields. In the unitary gauge, the interaction terms between
additional scalar particles and the Higgs boson are given by,
\begin{align}
{\cal L}_{\rm S2} \ni & - \l_3|H|^2|H_2|^2
- \l_4|H^\dagger H_2|^2 - \frac{\l_5}{2}[ (H_2^\dagger H)^2 + h.c. ] \nonumber\\
=&
-\l_3 |H^+|^2 \left( \frac{v+h}{\sqrt{2}} \right)^2
-\frac{\l_S}{2} s^2 \left( \frac{v+h}{\sqrt{2}} \right)^2
-\frac{\l_A}{2} a^2 \left( \frac{v+h}{\sqrt{2}} \right)^2,
\end{align}
where $\l_S \equiv \l_3 + \l_4 + \l_5 $ and $\l_A \equiv \l_3 + \l_4 -
\l_5 $ are effective couplings to the Higgs boson. 
Mass eigenvalues of them are given by,
\begin{align}
m_{H^+}^2 = m_2^2 + \frac{\l_3}{2} v^2,\qquad
m_{S^0}^2 = m_2^2 + \frac{\l_S}{2} v^2, \qquad
m_{A^0}^2 = m_2^2 + \frac{\l_A}{2} v^2. \label{eq:s2_mass}
\end{align}
$\l_A < \l_S$ and $m_{A^0} < m_{S^0}$ are satisfied because we take
$\l_5$ as real and positive.
Furthermore, if $\l_4 < \l_5$, $A^0$ becomes lighter than $H^+$.
In this situation, $A^0$ becomes the candidate of dark matter.
Conditions for the scalar potential to be bounded from below are given by \cite{Barbieri:2006dq},
\begin{align*}
\min[\l_A,\l_S,\l_3]=\l_A > -2\sqrt{\l_1\l_2},\qquad \l_2 > 0.
\end{align*}

\subsection{Triplet scalar / fermion model (Model S3, F3)}
If we add $SU(2)$ triplet scalar ($t$) / fermion ($\chi$) with $Y=0$, we can construct simple dark matter models\footnote{
One might think model F2, namely $SU(2)$ doublet Dirac fermion with
$Y=1/2$ is also simple possibility. 
However, a dark matter with non-zero hypercharge is severely constrained by the direct detection experiments \cite{Cirelli:2005uq}.
}~\cite{Araki:2011hm, Cirelli:2005uq}.
Here, we call them as model S3 and F3, respectively.
Model F3 is an effective theory of wino dark matter model \cite{Randall:1998uk, Giudice:1998xp}.
Lagrangian of each models are given by,
\begin{align}
{\cal L}_{\rm S3} &= {\cal L}_{\rm SM} + \frac{1}{2}(\q t)^2 - \frac{m_3^2}{2} t^2 - \frac{\l_{tH}}{2} t^2 |H|^2,\\
{\cal L}_{\rm F3} &= {\cal L}_{\rm SM} + i \psi_T \sla{\q}\psi_T - \left( \frac{M_T}{2} \psi_T\psi_T + h.c.\right).
\end{align}
In model S3, we can write the dark matter self interaction term $t^4$.
However, this self-interaction term does not affect our discussion, and thus we neglect it.

Model S3 has a neutral scalar $t^0$ and a charged scalar $t^+$,
and model F3 has a neutral Majorana fermion $\psi_T^0$ and a charged Dirac fermion $\psi_T^+$.
In both of the models, masses of the neutral particle and the charged particle are degenerated.
Mass splitting between them is generated by one-loop radiative correction \cite{Cirelli:2005uq}.
In model S3, for $m_{t^0}, m_{t^\pm} \gg m_W,~m_Z$,
\begin{align}
m_{t^+} - m_{t^0} &\simeq \frac{\a_2}{2}\left( m_W - c_W^2 m_Z \right) \simeq 166~{\rm MeV},
\end{align}
and, in model F3, for $m_{\chi^0}, m_{\chi^\pm} \gg m_W,~m_Z$,
\begin{align}
m_{\psi_T^+} - m_{\psi_T^0} &\simeq \frac{\a_2}{2}\left( m_W - c_W^2 m_Z \right) \simeq 166~{\rm MeV}.
\end{align}
Hence, in both of the models, charged particles becomes slightly heavier than neutral ones.
In such a situation, dark matter coannihilation becomes important to obtain a correct amount of thermal relic abundance,
and thus the mass of the dark matter tends to be large.
We can get correct relic abundance for $m_{t^0} \sim 2.5~\TEV$ in model S3, and $m_{\psi_T^0} \sim 2.7~\TEV$ in model F3 \cite{Cirelli:2007xd}.
It is difficult to discuss them in near future collider experiment. Hence, we do not discuss them in detail.

\subsection{Singlet-doublet fermion dark matter (Model F12)} \label{subsec:f12}
Here, we discuss singlet-doublet mixed fermion dark matter model \cite{Cohen:2011ec, Cheung:2013dua}.
We introduce three left-handed Weyl fermions;
SM singlet fermion $(\psi_S)$,
$SU(2)_L$ doublet fermion with $Y=-1/2$ $(\psi_D)$ and $Y=1/2$ $(\chi_D)$.
This matter content is vector-like, and this model is free from gauge anomaly.
Renormalizable interaction terms of dark matter sector are given by,
\begin{align}
{\cal L}_{\rm F12} 
&= 
-\frac{1}{2}m_S \psi_S \psi_S 
- m_D \psi_D \chi_D
+ y \tilde{H}^{\dagger} \psi_S \psi_D
+ y' H^{\dagger} \psi_S \chi_D + h.c.,
\end{align}
where $\tilde{H} = \epsilon H^{*}$ and $\epsilon$ is totally
antisymmetric tensor. We have four complex parameters in the dark matter sector.
Among them, three phases can be removed by a redefinition of $\psi_S$,
$\psi_D$ and $\chi_D$.
In this paper, we take a basis in which $m_S$, $m_D$ and $y$ are real
positive. 
Finally, we have the following five physical free parameters in this model,
\begin{align}
m_S,~ m_D,~ y,~|y'|,~ \theta\equiv{\rm arg}(y').
\end{align}

In this model, we have one charged Dirac fermion and three Majorana neutral fermions.
Mass of the charged fermion is $m_D$. Mass eigenstates of the neutral fermions are the mixture of $\psi_S$, $\psi_D^0$ and $\chi_D^0$.
Mass matrix of them is given by,
\begin{align}
{\cal L}_{\rm mass} &=
-\frac{1}{2}
\left( \begin{array}{ccc}
\psi_S & \psi_D^0 & \chi_D^0
\end{array} \right)
\left( \begin{array}{ccc}
m_S & y v/\sqrt{2} & -y' v/\sqrt{2} \\
y v/\sqrt{2} & 0 & -m_D \\
-y'v/\sqrt{2} & -m_D & 0
\end{array} \right)
\left( \begin{array}{c}
\psi_S \\ \psi_D^0 \\ \chi_D^0 \\
\end{array} \right).
\label{eq:f12_mass}
\end{align}
Mass eigenstates $f$'s can be written as linear combination of $\psi_S$, $\psi_D^0$ and $\chi_D^0$:
\begin{align}
\left( \begin{array}{c}
\psi_S\\ \psi_D^0\\ \chi_D^0
\end{array} \right)
=
\left( \begin{array}{ccc}
U_{11} & U_{12} & U_{13} \\
U_{21} & U_{22} & U_{23} \\
U_{31} & U_{32} & U_{33}
\end{array} \right)
\left( \begin{array}{c}
f_1^0\\ f_2^0\\ f_3^0
\end{array} \right),\label{eq:f12_mixing}
\end{align}
where $U$ is a unitary matrix. We define the following four-component Dirac and Majorana spinors:
\begin{align}
\Psi^+ \equiv \left(\begin{array}{c}
\chi_D^+ \\
\psi_D^{-\dagger} \\
\end{array}\right),\qquad
\Psi^0_i \equiv \left(\begin{array}{c}
f_i^0 \\
f_i^{0\dagger} \\
\end{array}\right),
\end{align}
Relevant interaction terms for the calculation of $S$, $T$ parameters and EDM are given by,
\begin{align}
{\cal L}_{\rm int.}
~=~& g \bar\Psi^+ \g^\m ({\cal C}_{L,i} P_L + {\cal C}_{R,i} P_R) \Psi^0_i W^+_\m 
+ g \bar\Psi^0_i \g^\m ({\cal C}_{L,i}^* P_L + {\cal C}_{R,i}^* P_R) \Psi^+ W^-_\m \nonumber\\
&+ \frac{g}{2c_W}\bar\Psi^0_i \g^\m ({\cal N}_{L,ij} P_L + {\cal N}_{R,ij} P_R) \Psi^0_j Z_\m
+ \frac{g}{c_W}\left(\frac{1}{2}-s_W^2 \right) \bar\Psi^+ \g^\m \Psi^+ Z_\m,
\end{align}
where ${\cal C}_{L,i}$, ${\cal C}_{R,i}$, ${\cal N}_{L,ij}$ and ${\cal N}_{R,ij}$ are determined by the mixing matrix $U$:
\begin{align}
{\cal C}_{L,i} = \frac{1}{\sqrt{2}}U_{3i},\qquad &{\cal C}_{R,i} = -\frac{1}{\sqrt{2}}U_{2i}^*,\qquad
{\cal N}_{L,ij} = -{\cal N}_{R,ji} = \frac{1}{2}(U_{3i}^* U_{3j} - U_{2i}^* U_{2j}).
\end{align}

Let us comment on symmetry in this model.
In the case of $\theta=0$ or $\pi$,
we can take all of the parameters in the dark matter sector as real by using redefinition of $\psi_S,~\psi_D$ and $\chi_D$,
{\it i.e.}, CP is conserved in dark matter sector.
On the other hand, in the cases of $\theta \neq 0, \pi$, the dark matter sector does violate CP symmetry.
The dark matter sector gives contribution to EDMs of the SM particles.
If $y=|y'|$, we have charge conjugation symmetry as $\psi_D \leftrightarrow \chi_D$.
In this case, dark matter-dark matter-$Z$ boson coupling vanishes.
Furthermore, $\psi_D$ and $\chi_D$ form an $SU(2)_R$ doublet and dark
matter sector has custodial symmetry in this case,
and the contribution of $Z_2$ odd particles to $T$-parameter vanishes at
the one-loop level.

\subsection{Doublet-triplet fermion dark matter (Model F23)} \label{subsec:f23}
Here, we discuss doublet-triplet mixed fermion dark matter model \cite{Dedes:2014hga}.
We introduce three left-handed Weyl fermions;
$SU(2)_L$ doublet fermion with $Y=-1/2$ $(\psi_D)$ and $Y=1/2$ $(\chi_D)$
and $SU(2)_L$ triplet fermion $(\psi_T)$ with $Y=0$.
This matter content is vector-like, and thus, it is free from gauge anomaly.
Renormalizable interaction terms in the dark matter sector are given by,
\begin{align}
{\cal L}_{\rm F23} &= -\frac{1}{2}m_S \psi_T \psi_T - m_D \psi_D \chi_D
+ y \tilde{H}^{\dagger} \psi_T \psi_D 
+ y' H^{\dagger} \psi_T \chi_D + h.c.
\end{align}
We have four complex parameters in the dark matter sector.
Among them, three phases can be removed by a redefinition of $\psi_D$, $\chi_D$ and $\psi_T$.
In this paper, we take a basis in which $m_D$, $m_T$ and $y$ are real
and positive.
In this basis, we have the following five physical free parameters,
\begin{align}
m_D,~ m_T,~ y,~ |y'|,~ \theta\equiv{\rm arg}(y').
\end{align}

In this model, there are two charged Dirac fermions and three Majorana
neutral fermions. 
Mass matrices of the fermions are given by,
\begin{align}
{\cal L}_{\rm mass} =&
-\frac{1}{2}
\left(
\begin{array}{ccc}
\psi_D^0 & \chi_D^0 & \psi_T^0
\end{array}
\right)
\left(
\begin{array}{ccc}
0 & -m_D & y v/\sqrt{2} \\
-m_D & 0 & -y'v/\sqrt{2}  \\
y v/\sqrt{2} & -y' v/\sqrt{2} & m_T
\end{array}
\right)
\left(
\begin{array}{c}
\psi_D^0 \\
\chi_D^0 \\
\psi_T^0 \\
\end{array}
\right) \nonumber\\
&- \left(
\begin{array}{cc}
\chi_D^+ & \psi_T^+
\end{array}
\right)
\left(
\begin{array}{cc}
m_D & y' v\\
y v& m_T
\end{array}
\right)
\left(
\begin{array}{c}
\psi_D^- \\
\psi_T^- \\
\end{array}
\right).
\end{align}
Mass eigenstates $f$'s can be written as linear combinations of $\psi_D$, $\chi_D$ and $\psi_T$:
\begin{align}
\left( \begin{array}{c}
\psi_D^0\\ \chi_D^0\\ \psi_T^0
\end{array} \right)
=
\left( \begin{array}{ccc}
U_{11}^0 & U_{12}^0 & U_{13}^0 \\
U_{21}^0 & U_{22}^0 & U_{23}^0 \\
U_{31}^0 & U_{32}^0 & U_{33}^0
\end{array} \right)
\left( \begin{array}{c}
f_1^0\\ f_2^0\\ f_3^0
\end{array} \right), \\
\left( \begin{array}{c}
\chi_D^+\\ \psi_T^+
\end{array} \right)
=
\left( \begin{array}{ccc}
U_{11}^+ & U_{12}^+ \\
U_{21}^+ & U_{22}^+ \\
\end{array} \right)
\left( \begin{array}{c}
f_1^+\\ f_2^+\\
\end{array} \right),\qquad
\left( \begin{array}{c}
\psi_D^-\\ \psi_T^-
\end{array} \right)
=
\left( \begin{array}{ccc}
U_{11}^- & U_{12}^- \\
U_{21}^- & U_{22}^- \\
\end{array} \right)
\left( \begin{array}{c}
f_1^-\\ f_2^-\\
\end{array} \right), \label{eq:f23_mixing}
\end{align}
where $U^0$, $U^+$ and $U^-$ are unitary matrices. We define the following four-component Dirac and Majorana spinors:
\begin{align}
\Psi^+_i \equiv \left(\begin{array}{c}
f^+_i \\
f^{-\dagger}_i \\
\end{array}\right),\qquad
\Psi^0_i \equiv \left(\begin{array}{c}
f_i^0 \\
f_i^{0\dagger} \\
\end{array}\right).
\end{align}

The situation is similar to the model F12 regarding of phases and
custodial symmetry.
In the case of $\theta=0$ and $\pi$,
we can take all the parameters in the dark matter sector as real by using redefinitions of $\psi_D,~\chi_D$ and $\psi_T$.
For $y=|y'|$, we have charge conjugation symmetry as $\psi_D
\leftrightarrow \chi_D$, which results in vanishing 
dark matter-dark matter-$Z$ boson coupling.
Due to the custodial symmetry at that point,
the contribution of $Z_2$ odd particles to the $T$-parameter vanishes at
the one-loop level.

\section{Phenomenology in each models}\label{sec:pheno}
In this section, we discuss phenomenological aspects of the dark matter models which are introduced in the previous section.
In the following analysis, we used FeynRules \cite{Alloul:2013bka} and micrOMEGAs \cite{Belanger:2013oya}
for the calculation of relic abundance of dark matter and Higgs invisible decay width.
We take the Higgs boson mass as 125~GeV~\cite{Aad:2014aba, CMS:2014ega}
throughout this paper.

\subsection{Model S1}\label{sec:S1}
In the model S1, there are only two parameters which are relevant to dark
matter physics, {\it i.e.}, the dark matter mass $m_{\rm DM}$ and the dark
matter-Higgs coupling $\l_{sH}$. 
By imposing the condition that the relic abundance explains the DM of
the Universe, $\lambda_{sH}$ is fixed as a function of $m_{\rm DM}$, and thus
$m_{\rm DM}$ is the only free parameter.
In the following we compute the direct detection cross section as
a function of the dark matter mass. If the dark matter mass is smaller
than a half of the Higgs boson mass, the Higgs boson can decay into two
dark matter particles. This contributes to the branching fraction of the invisible decay of the Higgs boson. We also discuss it here.

\subsubsection{Relic abundance and direct detection}
\begin{figure}[tb]
\centering
\includegraphics[width=0.49\hsize]{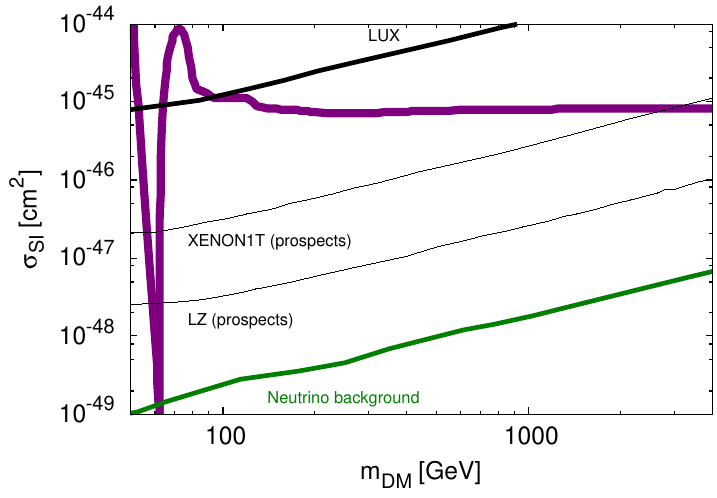}
\includegraphics[width=0.49\hsize]{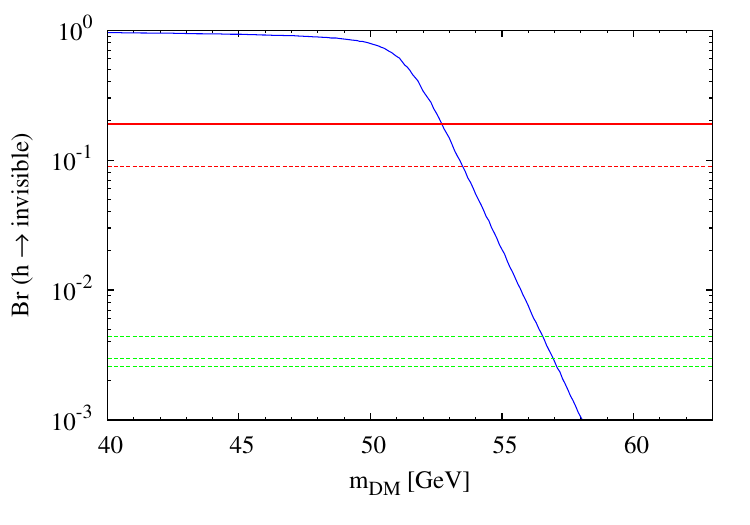}
\caption{ Present status of scalar dark matter S1 model.
\textbf{Left:} 
The wide mass range of the dark matter mass is shown. The purple line is
 the model prediction. The black bold solid line is current bound by the LUX experiment.
The green line shows the discovery limit which is caused by atmospheric and astrophysical neutrinos.
We also plot future prospect of XENON1T and LZ.
\textbf{Right:} The branching fraction of the Higgs invisible decay in S1 model.
The red solid line is the current bound.
The red dashed line is the future prospect of LHC at
 300~fb$^{-1}$.
The three green dashed lines are the future prospect of the ILC.
(250~GeV with 250 fb$^{-1}$,
500~GeV with 500 fb$^{-1}$,  
1~TeV with 1 ab$^{-1}$.
)
}\label{fig:mass-xsec-S1}
\end{figure}

Here, we show the constraint on the spin independent cross section
($\sigma_{\textrm{SI}}$) from 
the LUX experiment \cite{Akerib:2013tjd}.
The left panel of Fig.~\ref{fig:mass-xsec-S1}
shows $\sigma_{\rm SI}$ as a function of the dark matter mass $m_{\rm DM}$
where the correct dark matter abundance is imposed.
The mass region with $53~\GEV \lsim m_{\rm DM} \lsim 64~\GEV$ and $100~\GEV
\lsim m_{\rm DM}$ 
are allowed by the constraint from the LUX experiment.
We also show the future prospects of XENON1T and LZ \cite{Feng:2014uja},
and the dark matter discovery limit which is caused by atmospheric and astrophysical neutrinos \cite{Billard:2013qya}.

\subsubsection{Higgs invisible decay}
\begin{figure}[tb]
\centering
\includegraphics[width=0.60\hsize]{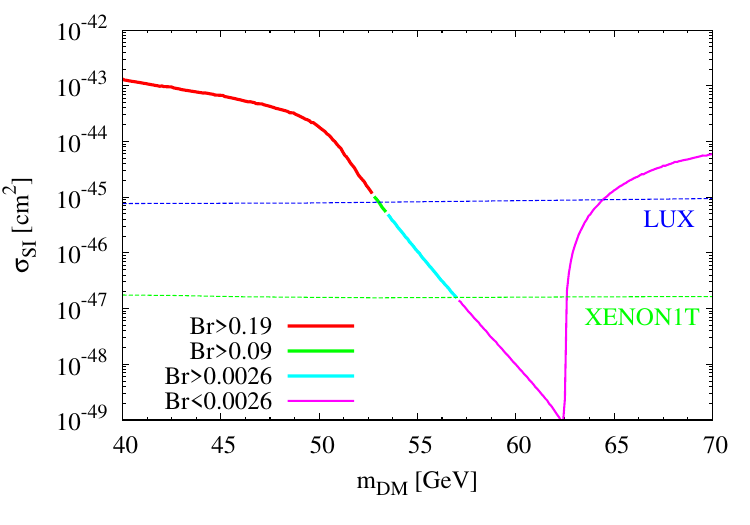}
\caption{
The dark matter mass around a half of the Higgs mass
 is shown. The line shows the parameter regions which can realize $\Omega_{\rm DM} h^2 = 0.12029$ \cite{Ade:2013zuv}.
The red, green, cyan, and purple solid lines are 
0.19 $<$ Br($h \to$ invisible), 
0.09 $<$ Br($h \to$ invisible) $<$ 0.19, 
0.0026 $<$ Br($h \to$ invisible) $<$ 0.09, 
and Br($h \to$ invisible) $<$ 0.0026,
respectively.
The green and blue dashed lines are the current bound by the LUX experiment and future
 prospect of XENON 1T.
}\label{fig:mass-invisible-S1}
\end{figure}

The right panel in Fig.~\ref{fig:mass-xsec-S1}
shows the branching fraction of the Higgs invisible decay as a function of the
dark matter mass $m_{\rm DM}$
while the requiring the correct dark matter abundance.
The current bound on the branching fraction of the invisible decay of the Higgs boson in model S1
is ${\rm Br}(h \to {\rm invisible}) <
0.19$~\cite{Belanger:2013xza}. This is shown with the red solid line in the
figure. Thus the lower bound on the dark matter mass is 53~GeV that
coincides the one from the LUX experiment. 
The LHC can reach Br = 0.09 with 300~fb$^{-1}$~\cite{Baer:2013cma}, and
the ILC can reach Br=0.0026 with $\sqrt{s}=1~{\rm TeV}$ and
1~ab$^{-1}$~\cite{Baer:2013cma}. These lines are shown in the
figure.

In the Fig.~\ref{fig:mass-invisible-S1},
we focus on the light DM mass region. The information on
the branching fraction of the Higgs invisible decay is also shown. 
We see that XENON1T will cover the ILC reach.
Therefore, in this model,
if XENON1T finds the DM signal in this region, the ILC should also find the
Higgs invisible decay.

\subsection{Model S2}\label{sec:S2}
In model S2,
the dark matter couplings to the SM particles are determined by $SU(2)_L \times U(1)_Y$ gauge
couplings and $\lambda_A$ which is defined in Sec.~\ref{subsec:s2}.
After fixing the size of $\lambda_A$ by the relic abundance, we discuss the spin independent
cross section at the direct detection experiments. We find that there are
two dark matter mass regions, $m_{\rm DM} \lsim 72~\GEV$ and $m_{\rm DM} \gsim 600~\GEV$.
We also discuss the mass bound on $m_{H^{+}}$ and $m_{S^{0}}$ from LEP2
and electroweak precision bounds.
Then we focus on the light dark matter mass region and discuss
the contribution of the dark matter to the branching fraction of the invisible decay, and the diphoton channel of the Higgs boson. We
will find the branching fraction of the invisible decay has similar behavior
as the S1 model, and the signal strength of $h\to \g\g$ is
$\sim$ 10\% smaller than the one in the SM.

\subsubsection{Relic abundance and direct detection}
\begin{figure}[t]
\centering
\includegraphics[width=0.48\hsize]{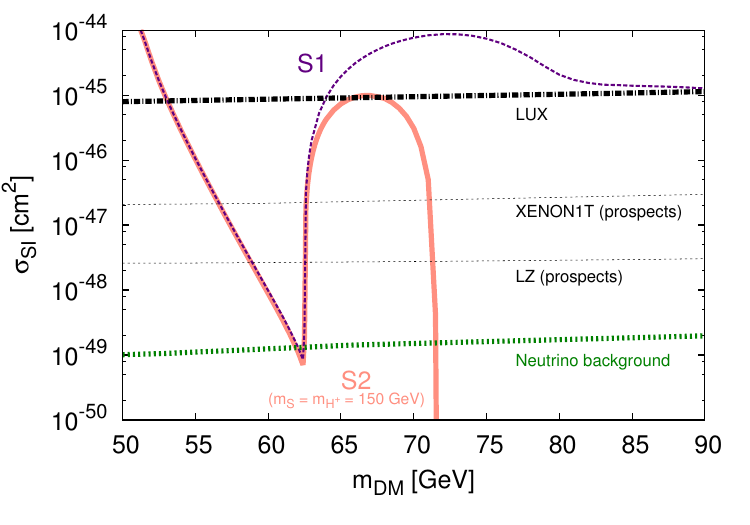}
\includegraphics[width=0.48\hsize]{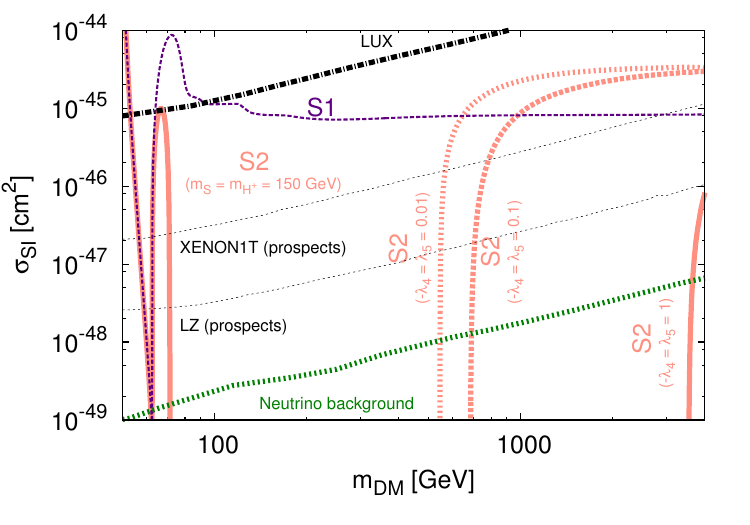}
\caption{
Present status of model S2.
Each line shows the parameter regions which can realize $\Omega_{\rm DM} h^2 = 0.12029$ \cite{Ade:2013zuv}.
Purple dotted line shows model S1, pink lines shows model S2.
In left figure, we take $m_{S^0}=m_{H^+}=150~\GEV$.
In right figure, we take $m_{S^0}=m_{H^+}=150~\GEV$ for leftmost pink line.
For other lines, we take $-\l_4 = \l_5$ as 0.01, 0.1 and 1 from left to right.
Black chain line shows the constraint from the LUX.
Green dotted lines show the discovery limit which is caused by atmospheric and astrophysical neutrinos \cite{Billard:2013qya}.
We also plot future prospect of XENON1T and LZ \cite{Feng:2014uja}.
}\label{fig:mass-xsec-S2}
\end{figure}

We have four free parameters, $m_{\rm DM}(=m_{A^0})$, $m_{S^0}$,
$m_{H^+}$ and $\l_A$. 
The value of $\l_A$ can be fixed by requiring the correct amount of
relic abundance. We have three mass parameters left.
In Fig.~\ref{fig:mass-xsec-S2},
we show the spin independent cross section $\s_{\rm SI}$ as a function
of dark matter mass $m_{\rm DM}$ 
in the dark matter S2 model
while requiring the correct dark matter abundance.

The parameter space which gives the correct amount of the dark matter is splitted into two regions;
the light mass region $m_{\rm DM} \lsim 72~\GEV$ and the heavy mass region $m_{\rm DM} \gsim 500~\GEV$.
In the light mass region, the Higgs boson $s$-channel exchange diagrams give dominant contribution to the dark matter annihilation cross section.
If $m_{S^0}-m_{A^0}$ and $m_{H^+}-m_{A^0}$ are large enough,
as we can see from Fig.~\ref{fig:mass-xsec-S2}, the S2 dark matter has
very similar behaviour to S1 dark matter at the tree level calculation.
However, for $m_{\rm DM} \sim m_h/2$ ({\it i.e.}, small $\l_A$), it is pointed out that one-loop radiative corrections give significant modification on
the spin independent cross section for model S2 dark matter \cite{Klasen:2013btp},
because it is charged under the electroweak gauge group.
In the figure, the tree level cross section is shown.
For $m_{\rm DM} > m_W$, $A^0 A^0\to W^+ W^-$ channel opens and the annihilation cross section becomes large.
Therefore unlike the model S1, the abundance of S2 dark matter becomes too small to explain $\Omega_{\rm DM}h^2$ for $m_{\rm DM} \gsim m_W$.
For $m_{\textrm{DM}} \gtrsim 500$~GeV, a viable region reappears. 
The relic abundance in the heavy mass region is very sensitive to
mass splittings between dark matter and heavier particles $S^0$ and
$H^{\pm}$.

\subsubsection{Direct search}\label{sec:direct-search-S2}
Now we determined two parameters, $m_{\rm DM}$ and $\lambda_A$ by imposing the
correct relic density. The remaining
parameters are $m_{H^{+}}$ and $m_{S^{0}}$. 

\begin{figure}[t]
\centering
\includegraphics[width=0.49\hsize]{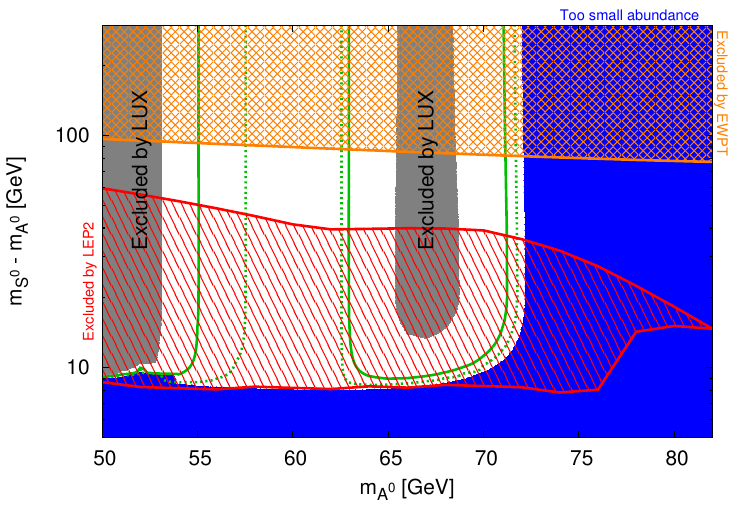}
\includegraphics[width=0.49\hsize]{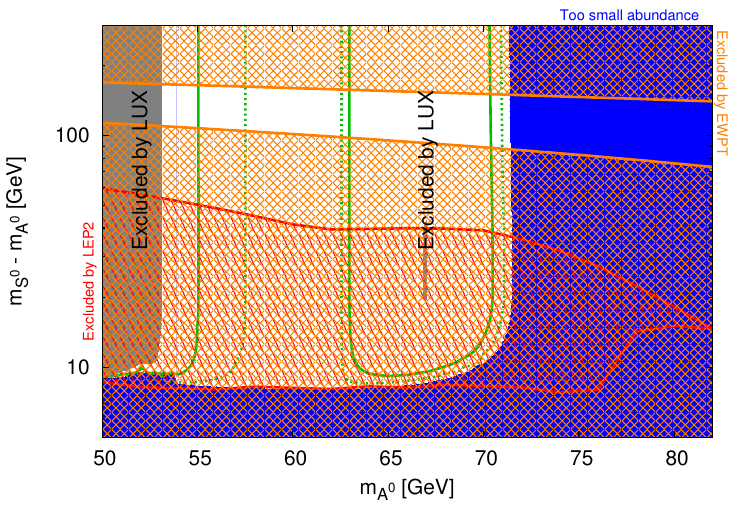}
\includegraphics[width=0.49\hsize]{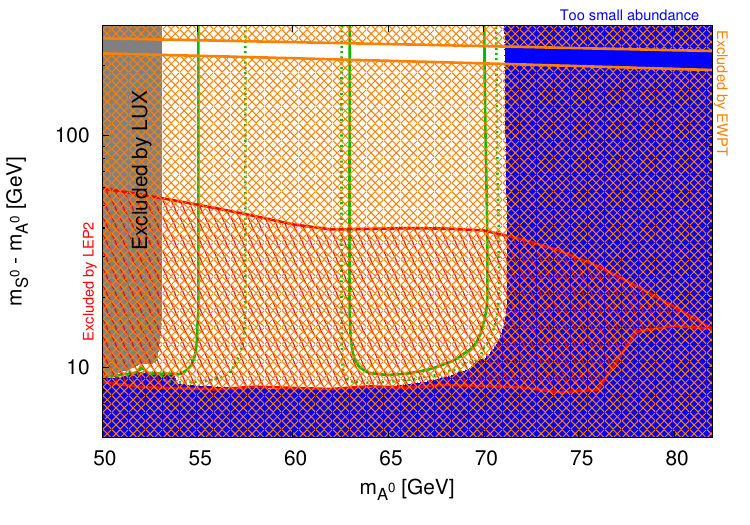}
\caption{
LEP2 constraint in $m_{A^0}$--$(m_{S^0}-m_{A^0})$ plane.
In the blue regions, thermal relic abundance becomes too small.
In this figure, we take $m_{H^+} = 120~\GEV$ (upper-left), $m_{H^+} =
 200~\GEV$ (upper-right), and $m_{H^+} = 300~\GEV$ (lower).
The red meshed regions are excluded by LEP2 experiment \cite{Lundstrom:2008ai}.
The gray regions are excluded by the LUX experiment. The yellow regions are excluded by $S$ and $T$ parameters.
The green solid and the dotted lines show $\s_{\rm SI} = 10^{-46}$ and $10^{-47}~{\rm cm}^2$, respectively.
}\label{fig:s2atlep2}
\end{figure}

At $e^+ e^-$ colliders, $H^+$ can be produced by a process $e^+ e^-\to (Z/\g)^* \to H^+ H^-$.
The LEP2 experiment gives the lower bound of the mass of $H^\pm$ to be 70--90 GeV \cite{Pierce:2007ut}.
Also, neutral scalar bosons can be produced by a process $e^+ e^- \to Z^* \to A^0 S^0$ \cite{Lundstrom:2008ai}. See Fig.~\ref{fig:s2atlep2}.
For the dark matter lighter than the $W$ boson, we find that a viable parameter
regions opens for  $m_{S^0} - m_{A^0} \gsim 40~\GEV$, and a small window around
$m_{S^0} - m_{A^0} \simeq 8~\GEV$. 

For $65~\GEV < m_{\text{DM}} < 70~\GEV$, we see that the exclusion by
the spin-independent cross section is sensitive to the charged Higgs
mass. In this region, we need to take into account $A^0 A^0 \to WW^{*}$
process in the relic density.
The diagram exchanging the charged Higgs in $t$-channel contributes to
this process, and
it is destructive with the same process with the Higgs boson
in the s-channel. Therefore, the heavier charged Higgs requires the
smaller Higgs coupling to the DM to reproduce the correct relic
abundance, and thus the spin-independent cross section is also smaller when
the charged Higgs is heavier and heavier.
Note that the spin-independent cross section in this region
is on the edge of the exclusion limit as we can see from Fig.~\ref{fig:mass-xsec-S2}.  

At the LHC, the model S2 can be probed by searching for 
dilepton and missing energy signal \cite{Dolle:2009ft} and 
trilepton and missing energy signal \cite{Miao:2010rg}.
For $40~\GEV \lsim m_{\rm DM} \lsim 72~\GEV$,
this search has a sensitivity in the parameter region with $m_{H^+,S^0} \simeq 100$--$180~\GEV$.

\subsubsection{$S$ and $T$ parameters}\label{sec:ST-S2}
\begin{figure}[t]
\centering
\includegraphics[width=0.45\hsize]{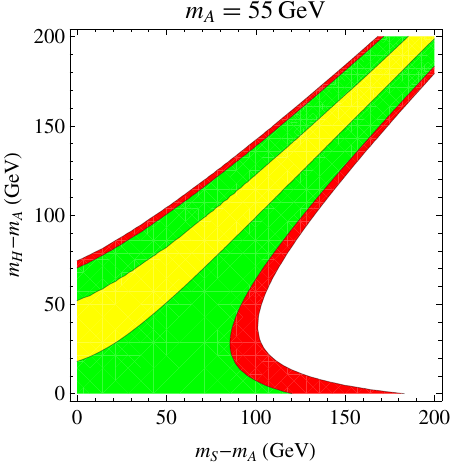}
\caption{
The constraint from the $S$ and $T$ parameters for $m_{\rm DM} = 55$~GeV.
The yellow, green, and red regions are allowed at 31.7\% C.L., 90\% C.L., and 95\% C.L., respectively.
}\label{fig:ST_S2}
\end{figure}

In this section, we discuss electroweak precision measurement.
The gauge boson two-point functions are given as,
\begin{align}
\parbox[c][2cm][c]{5cm}{ \includegraphics[width=5cm]{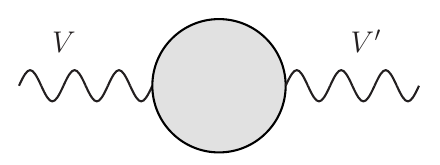} } &= i\Pi_{VV'}(p^2) g^{\m\n} + i\D_{VV'} p^\m p^\n.
\end{align}
By using this, the Peskin-Takeuchi parameters \cite{Peskin:1991sw} are
defined as, 
\begin{align}
S &= \frac{4s^2c^2}{\a}\left( \frac{\Pi_{ZZ}(m_Z^2)-\Pi_{ZZ}(0)}{m_Z^2} - \frac{c^2-s^2}{sc} \frac{\Pi_{Z\g}(m_Z^2)}{m_Z^2} - \frac{\Pi_{\g\g}(m_Z^2)}{m_Z^2}  \right), \label{eq:peskintakeuchi_s}\\
T &= \frac{1}{\a}\left( \frac{\Pi_{WW}(0)}{m_W^2} - \frac{\Pi_{ZZ}(0)}{m_Z^2} \right). \label{eq:peskintakeuchi_t}
\end{align}
The contributions to $\Pi_{VV'}$'s from the dark matter sector are given by,
\begin{align}
\Pi_{WW} 
=& 
\frac{g^2}{16\pi^2} \tilde B_{22}(m_{H^+},m_s) 
+ 
\frac{g^2}{16\pi^2} \tilde B_{22}(m_{H^+},m_a)
, \\
\Pi_{ZZ} 
=& 
\frac{1}{16\pi^2}
\frac{g^2}{c^2}
\left(
\tilde B_{22}(m_s, m_a)
+
(c^2 - s^2)^2
\tilde B_{22}(m_{H^{+}}, m_{H^+})
\right)
,\\
\Pi_{Z\g} 
=& 
\frac{eg}{c}\frac{c^2-s^2}{8\pi^2} \tilde B_{22}(m_{H^+},m_{H^+})
,\\
\Pi_{\g\g} 
=& 
\frac{e^2}{4\pi^2} \tilde B_{22}(m_{H^+},m_{H^+})
.
\end{align}
The definitions of $\tilde B_{22}$ are given in the Appendix \ref{sec:oblique}\footnote{
We have checked that our formulae are consistent with
Ref.~\cite{Barbieri:2006dq} in the limit of $m_{S^0},~m_{A^0},~m_{H^+} \gg m_Z$.
}.
We show the numerical result of the constraint for $m_{\rm DM} = 55$~GeV
in Fig.~\ref{fig:ST_S2}. We find that large mass
difference between the dark matter and other $Z_2$ odd particles are
disfavored except for $m_{S^0} \sim m_{H^+}$ case.
Note that when $m_{H^{+}} = m_{S^{0}}$, the custodial symmetry appears
in the $Z_2$ odd sector, and thus the $T$ parameter becomes zero at the
one-loop level. 
The constraints are superimposed in Fig.~\ref{fig:s2atlep2}.

\subsubsection{Higgs invisible decay}
Since there are viable parameter regions for $m_{A^0} < m_h/2$,
we have a chance to observe the invisible decay of the Higgs boson.
From the discussion on the LEP2 bound in Sec.~\ref{sec:direct-search-S2}
and the discussion on the electroweak
precision bound in Sec.~\ref{sec:ST-S2}, it is natural to expect that
$m_{S^{0}} \simeq m_{H^{+}} \gtrsim m_{\rm DM} + 40$~GeV. 
Then the S2 dark matter has very similar behaviour to the S1 dark matter as
we can see from Fig.~\ref{fig:mass-xsec-S2}. We conclude that the branching fraction of the Higgs invisible
decay as a function of the dark matter mass in this model behaves
same as in the S1 model shown in the right panel in Fig.~\ref{fig:mass-xsec-S1}.

\subsubsection{Higgs diphoton decay signal}
\begin{figure}[t]
\centering
\includegraphics[width=0.5\hsize]{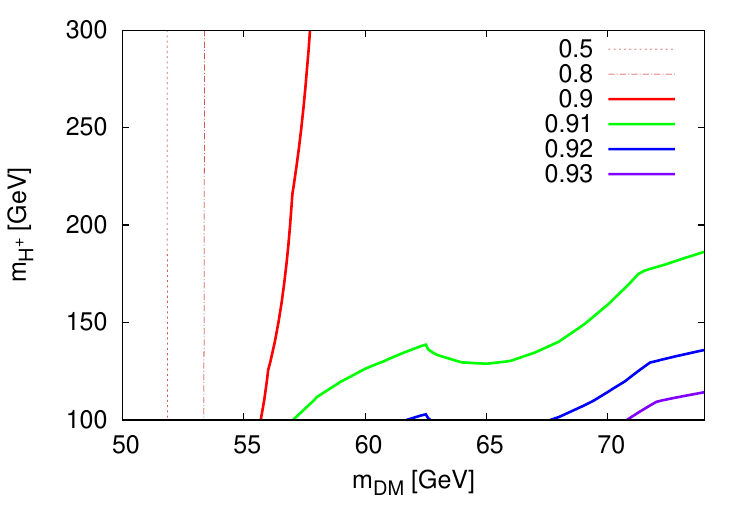}
\caption{
Diphoton signal strength $\m \equiv {\rm Br}(h\to2\g) / {\rm Br}(h\to2\g;~{\rm SM})$ in $m_{\rm DM}$--$m_{H+}$ plane.
In this figure, we take $m_{S^0} = m_{\rm DM} + 100~\GEV$.
}\label{fig:diphoton_s2}
\end{figure}

In model S2, loop diagrams including a charged scalar $H^+$ modify the branching fraction of the Higgs boson into two photons.
Its decay width is given by \cite{Arhrib:2012ia, Djouadi:2005gi},
\begin{align}
\G(h\to 2\g)
&= \frac{G_F \a^2 m_h^3}{128\sqrt{2} \pi^3}\left|
A_{\rm SM} + \frac{\l_3 v^2}{2m_{H^+}^2} A_0\left(\frac{m_h^2}{4m_{H^+}^2}\right) \label{eq:diphoton_s2}
\right|^2,
\end{align}
where the second term in the absolute value is the
contribution from $H^+$, 
and $A_{\rm SM}$ is the contribution from the SM particles, which is given by,
\begin{align}
A_{\rm SM} = 
A_1\left( \frac{m_h^2}{4m_W^2} \right) + \sum_f N_C Q_f^2 A_{1/2}^H\left( \frac{m_h^2}{4m_f^2} \right),
\end{align}
and its numerical value is $A_{\rm SM} \simeq -6.45$ for $m_h = 125~\GEV$, $m_W = 80.4~\GEV$ and $m_t = 173~\GEV$.
The definition of the function $A$'s are given in the Appendix \ref{sec:diphotonloopfunc}.
We can expect that the charged Higgs contribution does not decouple 
even if the charged Higgs mass is much larger than the electroweak scale
as long as
the mass difference between the dark matter and the charged Higgs mass
is kept large. Because the mass differences among the $Z_2$ odd particles
imply the sizable value of couplings of the $Z_2$ odd particles to the Higgs boson, namely sizable
$\lambda_{3, 4, 5}$, the charged Higgs coupling to the Higgs boson
remains even if its mass is quite large.
We can confirm this expectation from Eq.~(\ref{eq:diphoton_s2}).
By using Eq.~(\ref{eq:s2_mass}),
$\l_3$ can be written by $m_{H^+}$, $m_{A^0}$ and $\l_A$,
\begin{align}
\frac{\l_3 v^2}{2m_{H^+}^2} = 1 - \frac{m_{A^0}^2}{m_{H^+}^2} + \frac{\l_A v^2}{2m_{H^+}^2}.
\end{align}
For $x\ll 1$, $A_0(x) \simeq 1/3 + 8x/45 + \cdots$.
Thus, as long as we consider light dark matter $A^0$, even if the charged scalar
is relatively heavy, the charged Higgs contribution remains and
its asymptotic behavior is 
$\lambda_3 v^2 /(2 m_{H^{+}}^2) A_0 \to 1/3$   $(m_{H^{+}} \to \infty)$.

We show how the diphoton branching fraction is modified in Fig.~\ref{fig:diphoton_s2}.
We find that the branching fraction to the diphoton channel deviates from the
standard model around 10 \%.
Sensitivity to the diphoton signal strength is around 10~\% at the LHC 14 TeV $300~{\rm fb}^{-1}$,
 and it reaches around 5~\% at the ILC \cite{Peskin:2012we}.
We can conclude that model S2 can be probed at the ILC in the case of $m_{\rm DM} \lsim 72~\GEV$.

\subsection{Model F12}\label{sec:F12}
In this section, we discuss phenomenological aspects of model F12.
One of the features of this model is a CP-violating phase,
and as we will see, it has important effects on dark matter phenomenology.

\subsubsection{Relic abundance and direct detection}
\begin{figure}[t]
\centering
\includegraphics[width=0.48\hsize]{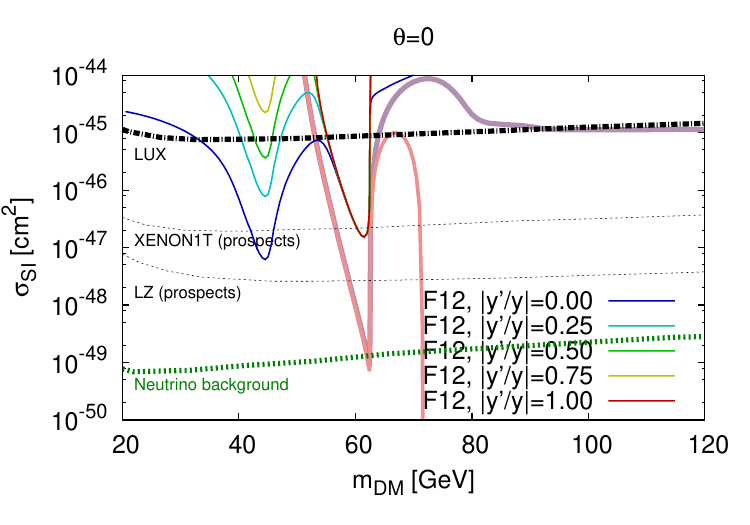}
\includegraphics[width=0.48\hsize]{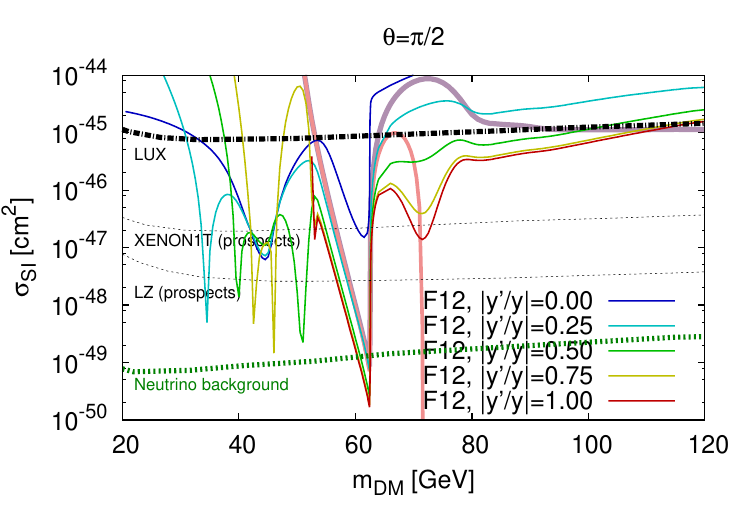}\\
\includegraphics[width=0.48\hsize]{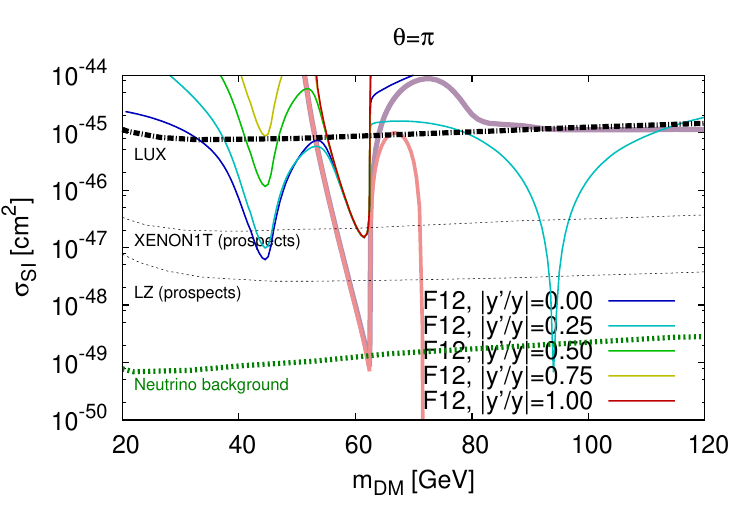}
\includegraphics[width=0.48\hsize]{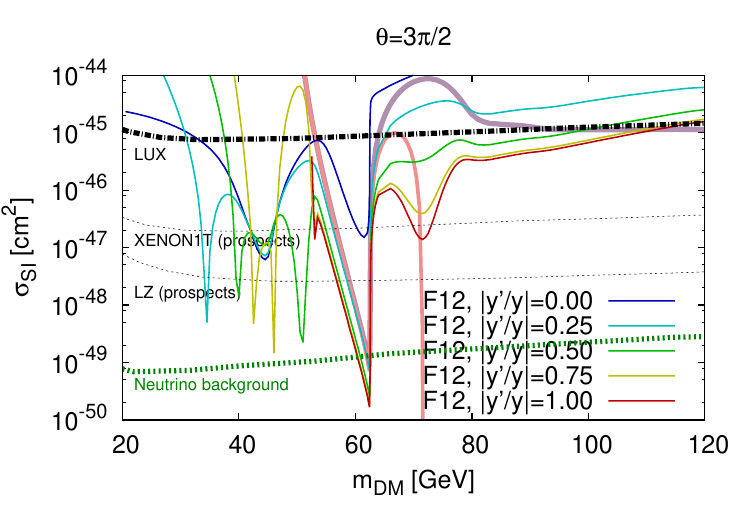}
\caption{
The spin independent cross section $\s_{\rm SI}$ of model F12.
Blue, cyan, green, yellow, and red lines show $|y'/y|=0,~0.25,~0.5,~0.75$ and 1 respectively.
We take $m_D = 200~\GEV$ in these figures.
For model F12, we take $\theta = 0$ (upper left), $\pi/2$ (upper right), $\pi$ (lower left) and $3\pi/2$ (lower right).
We also show $\s_{\rm SI}$ for model S1 (gray line) and model S2 with $m_{S^0} = m_{H^+} = m_{A^0} + 100~\GEV$ for references.
}\label{fig:mass-xsec-fermion}
\end{figure}

\begin{figure}[t]
\centering
\includegraphics[width=0.48\hsize]{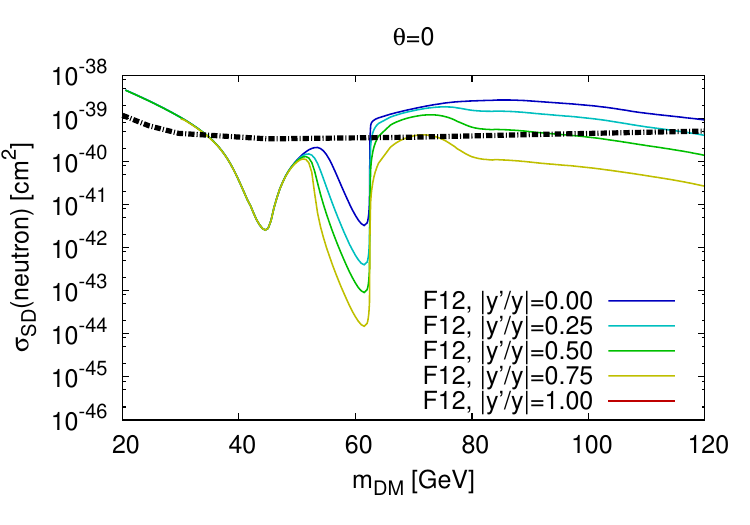}
\includegraphics[width=0.48\hsize]{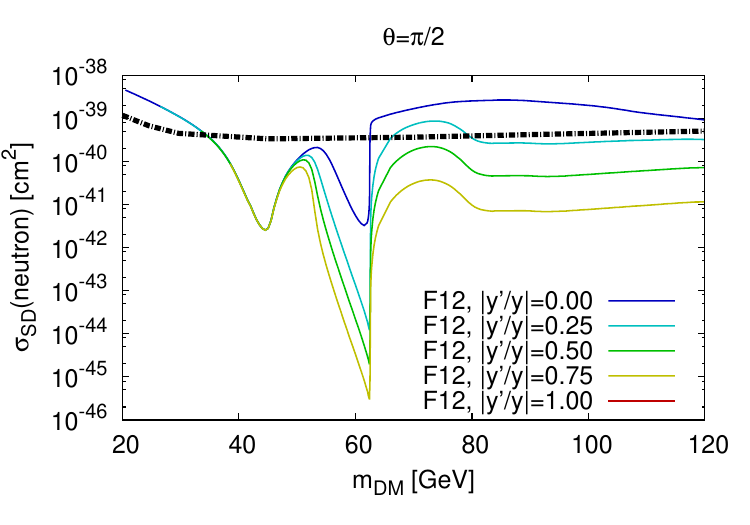}\\
\includegraphics[width=0.48\hsize]{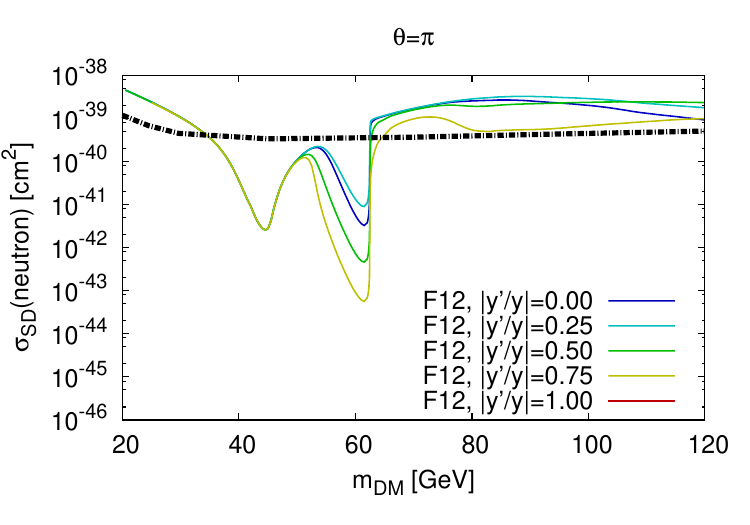}
\includegraphics[width=0.48\hsize]{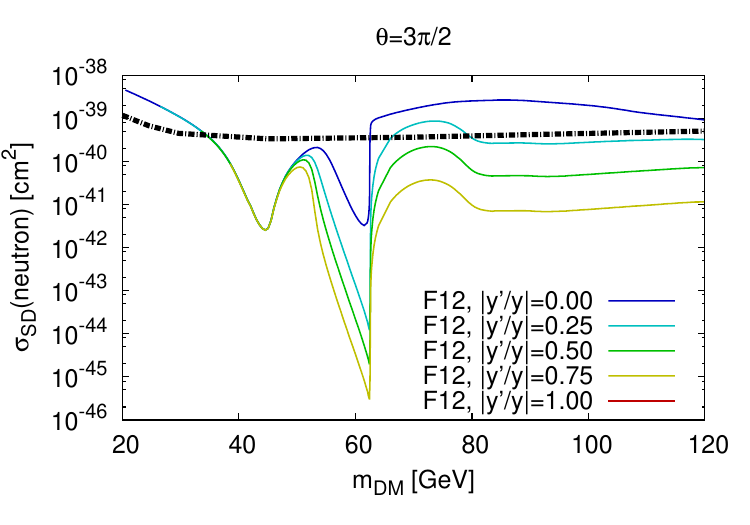}
\caption{
The spin dependent cross section $\s_{\rm SD}$ of model F12.
Blue, cyan, green, yellow, and red lines show $|y'/y|=0,~0.25,~0.5,~0.75$ and 1 respectively.
We take $m_D = 200~\GEV$ in these figures.
For model F12, we take $\theta = 0$ (upper left), $\pi/2$ (upper right), $\pi$ (lower left) and $3\pi/2$ (lower right).
We also show $\s_{\rm SI}$ for model S1 (gray line) and model S2 with $m_{S^0} = m_{H^+} = 150~\GEV$ for references.
Black lines shows the constraint on the spin-dependent cross section of neutron-WIMP from XENON 100 \cite{Beltrame:2013bba}.
}\label{fig:F12_xsec_SD}
\end{figure}

\begin{figure}[p]
\centering
\includegraphics[width=0.48\hsize]{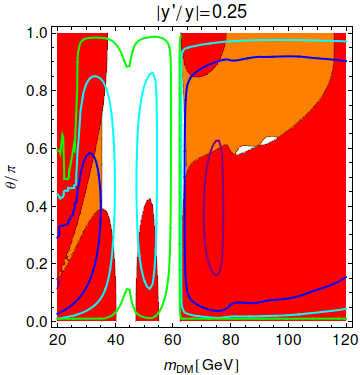}
\includegraphics[width=0.48\hsize]{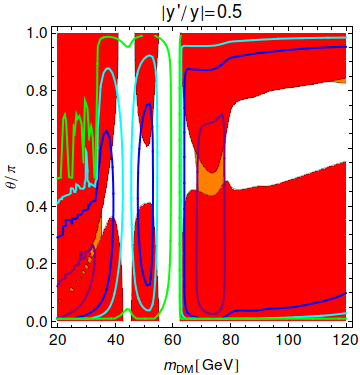} \\[1cm]
\includegraphics[width=0.48\hsize]{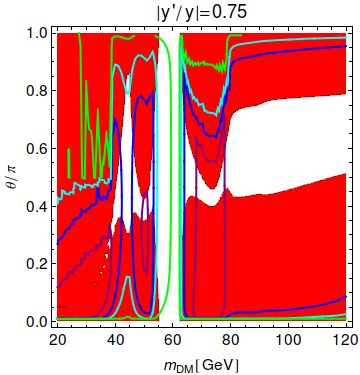}
\includegraphics[width=0.48\hsize]{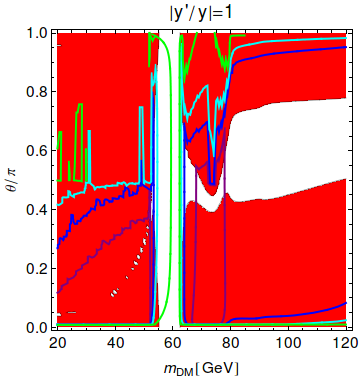}
\caption{
The electron EDM in $m_{\rm DM}$--$\theta$ plane.
We take $|y'/y| = 0.25, 0.5, 0.75, 1$ in each figure, and $m_D = 200~\GEV$ in all of the figures.
At each point, we set overall size of the Yukawa couplings $y$ and $y'$ to realize $\Omega_{\rm DM} h^2 = 0.12$.
Green, cyan, blue, and purple lines shows 
$|d_e| = 10^{-30}, 10^{-29},~3 \times 10^{-29}$ and $9\times
 10^{-29}~e{\rm cm}$,  respectively.
The red regions are excluded by the constraint on the spin-independent cross section by the LUX experiment \cite{Akerib:2013tjd}.
The orange regions are not excluded by the LUX but are excluded by the
 constraint on the spin-dependent cross section by the XENON100 experiment \cite{Beltrame:2013bba}.
}\label{fig:mass-theta}
\end{figure}

We show spin-independent direct detection cross section for model F12 in
Fig.~\ref{fig:mass-xsec-fermion}. Here
we consider the case that the model gives the correct dark matter abundance.
Similar to model S1 and S2, direct detection gives severe constraint on F12.
In the case of the dark matter mass around $m_h/2$ and $m_Z/2$, the spin independent cross section becomes small.
This is because, in these dark matter mass regions, diagrams with Higgs
boson and $Z$ boson in $s$-channel give 
the dominant contribution to the annihilation cross section, which
requires small Higgs/$Z$ boson coupling to the DM.

In addition to this structure, the direct detection cross section shows
complicated structures when we turn on the phase of the Yukawa coupling,
$\theta$. See, for example, 30~GeV $\lesssim m_{\textrm{DM}}
\lesssim$ 50~GeV region in the right panels in
Fig.~\ref{fig:mass-xsec-fermion}. We can understand this behavior as follows.
The mass term and the interaction terms of the dark matter with the SM particles are written as,
\begin{align}
{\cal L} 
\ni 
-\frac{m_{\rm DM}}{2} \bar\Psi_1^0 \Psi_1^0 
+ y_S h \bar\Psi_1^0 \Psi_1^0 
+ iy_P h \bar\Psi_1^0 \g^5 \Psi_1^0
+ i c_Z Z_{\mu} \bar\Psi_1^0 \g^{\mu} \g^5 \Psi_1^0.
\end{align}
Here, $y_S$, $y_P$, and $c_Z$ are calculated from $\l$, $\l'$ and
the unitary matrix $U$ which is defined in Eq.~(\ref{eq:f12_mixing}). 
Although all couplings ($y_S$, $y_P$, and $c_Z$) contribute to the
annihilation cross section, 
only $y_S$ contributes to the spin independent cross section.
This means that the spin independent cross section becomes zero while the
correct relic abundance can be explained when $y_S =0$, $y_P \neq 0$, and $c_Z \neq 0$.
We found $y_S=0$ when the following condition is satisfied:
\begin{align}
&
m_{\rm DM} = m_S = - m_D \sin 2\phi \cos\theta
\quad
(\theta = 0, \pi)
,
\nonumber\\
&
m_{\rm DM}^2 = \frac{ m_S^2 m_D^2 \sin^2 2\phi \sin^2\theta}{m_S^2 + m_D^2 \sin^2
 2\phi + 2m_S m_D \sin2\phi \cos\theta}
\quad
(\theta \neq 0, \pi)
,
\end{align}
where $\tan\phi = |y/y'|$.
Here we take $m_S>0$ and $m_D >0$ by using the freedom of the field
redefinition, thus $\theta =0$ can not satisfy this condition.
When this condition is satisfied, we have a sizable annihilation cross section and small spin independent cross section.
Such a parameter region is called as ``blind spot''~\cite{Cheung:2013dua}.

We also show the spin-dependent direct detection cross section for model F12
given the correct dark matter abundance in Fig.~\ref{fig:F12_xsec_SD}.  
By comparing with Fig.~\ref{fig:mass-xsec-fermion}, we find that the
spin-dependent cross section gives weaker bound than the spin-independent
cross section in wide region. Exception is the blind spots. In the blind
spots, the dark matter couplings to the $Z$ boson and to the Higgs boson
with $\gamma^5$ are needed to reproduce the relic abundance, so the
coupling to the $Z$ boson can be large enough to make the spin-dependent
cross section larger than the current bound. This is crucial in the
blind spots for $\theta=\pi$ case because the dark matter couplings to
the Higgs boson completely vanish in this case, and thus the dark matter
coupling to the $Z$ boson must be sizable. We see this feature
in the bottom-left panel in Figs.~\ref{fig:mass-xsec-fermion} and
\ref{fig:F12_xsec_SD}. In these panels, $\theta=\pi$ and there is a
blind spot for $m_{\rm DM} \simeq 90$~GeV and $|y'/y| = 0.25$, and we find
this region is already excluded by the bound on the spin-dependent cross
section. 

%
Non-vanishing CP phase significantly enlarges the viable mass range of the dark matter by having $y_S$ and $y_P$ simultaneously.
We show in Fig.~\ref{fig:mass-theta} the contour of the spin-independent cross section for various $|y' / y|$ ratios in the $m_{\rm DM}$--$ \theta$ plane.
As we will see later, such a CP phase induces EDM of the electron, and thus wide range of parameters can be covered by future EDM measurements.

\subsubsection{Higgs invisible decay}
\begin{figure}[t]
\centering
\includegraphics[width=0.48\hsize]{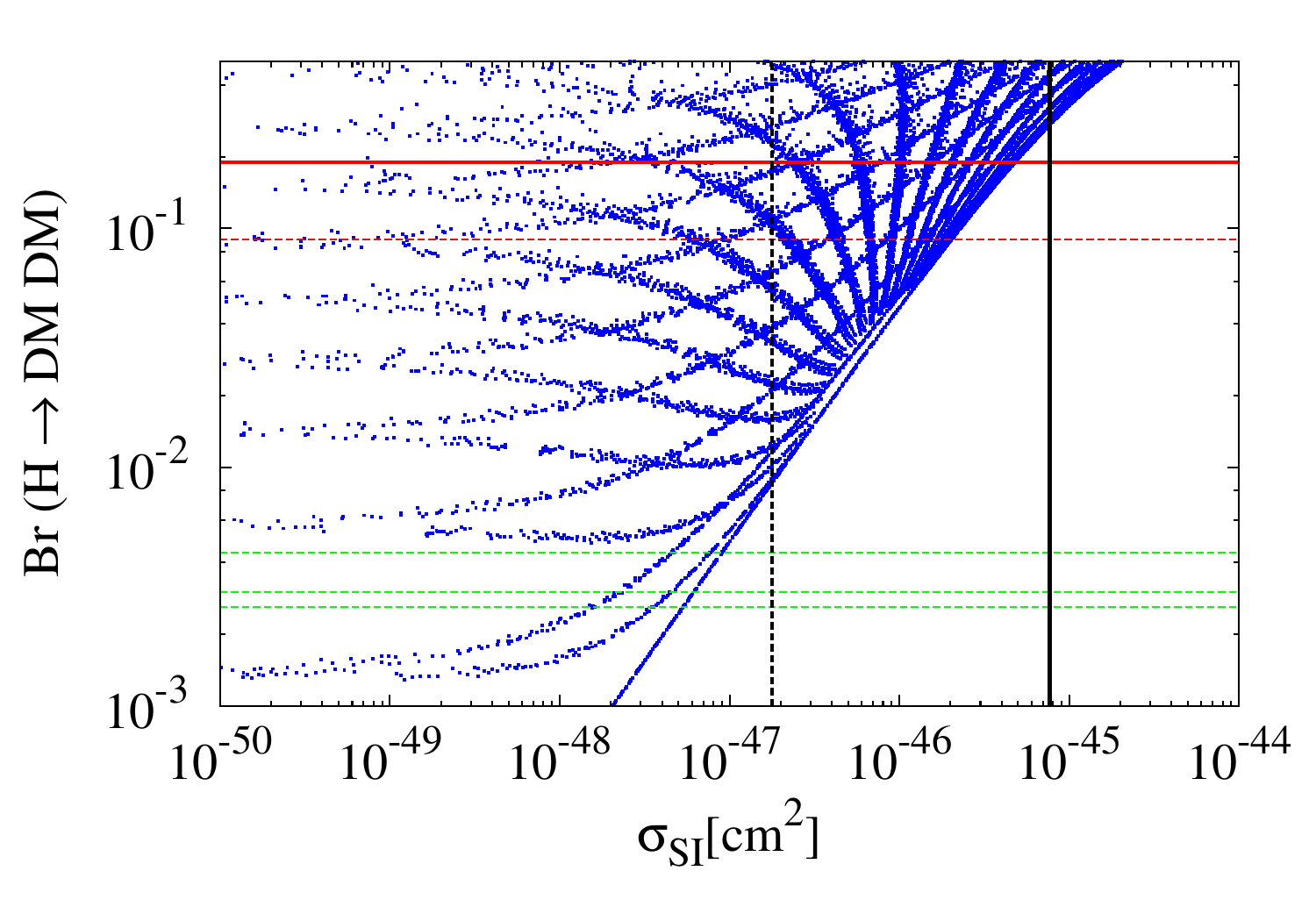}
\includegraphics[width=0.48\hsize]{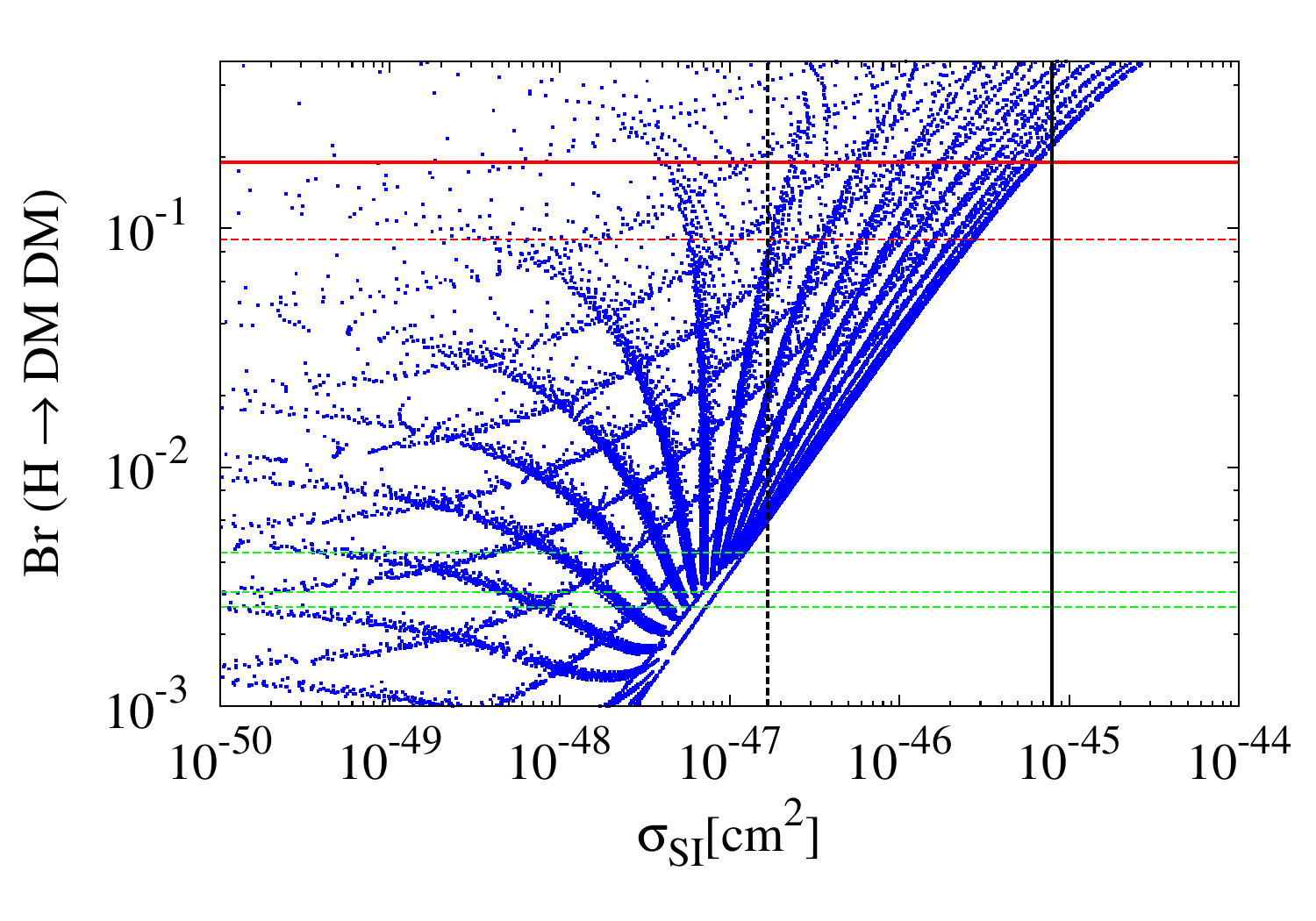}
\caption{
The spin independent cross section versus ${\rm Br}(h\to 2{\rm DM})$ for
 $m_{\rm DM} = 40~\GEV$ (left panel) and 
 $m_{\rm DM} = 45~\GEV$ (right panel) 
in model F12. 
See text for the parameter we used.
The red solid line is the current bound~\cite{Belanger:2013xza}.
The red dashed line is the future prospect of LHC at 300~fb$^{-1}$~\cite{Baer:2013cma}.
The three green dashed lines are the future prospect of the ILC.
(250~GeV with 250 fb$^{-1}$,
500~GeV with 500 fb$^{-1}$,  
1~TeV with 1 ab$^{-1}$~\cite{Baer:2013cma}.
)
The black solid line is the current 
bound by the LUX experiment. The black dashed line is the future prospect by XENON1T. 
}\label{fig:sigma_vs_invisible}
\end{figure}

If the mass of the dark matter is smaller than a half of the Higgs boson mass,
the Higgs invisible decay channel opens and we can use it as a probe of the dark matter sector.
In Fig.~\ref{fig:sigma_vs_invisible}, we show the branching fraction of the invisible decay of model F12.
We calculate the partial decay width for the invisible decay $\G_{\rm inv.}$ by using micrOMEGAs.
The decay width of the Higgs boson in the SM is calculated as $\G_{\rm SM} = 4.41 \times 10^{-3}~\GEV$ for $m_h= 125~\GEV$ by using HDECAY \cite{Djouadi:1997yw}.
The branching fraction of the invisible decay is given by $\G_{\rm inv.}/ ( \G_{\rm inv.} + \G_{\rm SM})$.
In Fig.~\ref{fig:sigma_vs_invisible},
we vary  three parameters, ($\theta$, $|y'/y|$, $m_2$), and their
values are
\begin{align}
 \theta/\pi =& 0, 0.05, 0.01, \cdots, 1.00, \\
 |y'/y|^{-1} =& 1, 1.5, 2.0, \cdots, 20.0, \\
 m_2 =& 150, 160, \cdots, 500.
\end{align}
Other parameters are fixed by the dark matter mass and the relic
abundance. 
We find that the smaller spin independent cross section means the
smaller branching fraction of the Higgs invisible decay . We also find that the ILC can
detect the signal of this model by searching for the Higgs invisible decay
even if the XENON1T experiment does not find any dark matter signals. This is
different feature of this model from model S1 and S2.
Again, by having $y_S$ and $y_P$ coupling simultaneously, the invisible width can be large even if the spin-independent cross section is small.

\subsubsection{$S$ and $T$ parameters}
\begin{figure}[t]
\includegraphics[width=5cm]{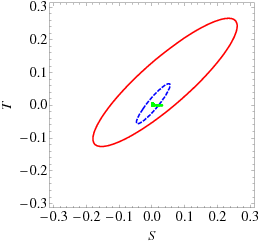}
\includegraphics[width=5cm]{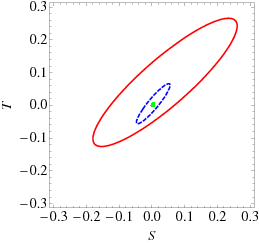}
\includegraphics[width=5cm]{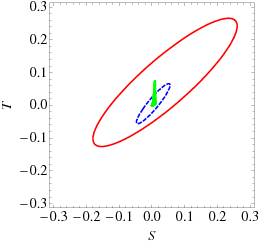}
\includegraphics[width=5cm]{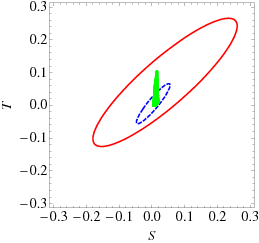}
\includegraphics[width=5cm]{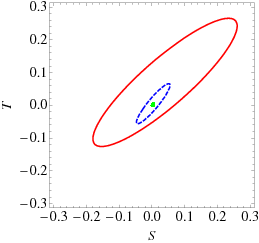}
\caption{The $S$ and $T$ parameters for 
$m_{\rm DM}=40$~GeV (upper-left),
$m_{\rm DM}=45$~GeV (upper-middle),
$m_{\rm DM}=60$~GeV (upper-right),
$m_{\rm DM}=70$~GeV (lower-left),
$m_{\rm DM}=200$~GeV (lower-right).
Red line is current bound at 95\% C.L.~\cite{Beringer:1900zz}.
Blue dashed line is GFITTER's future prospect at the ILC~\cite{Baak:2014ora}.
The green dots are consistent points with the current direct search
 result by the LUX experiment.
}
\label{fig:st_f12}
\end{figure}

The contributions to $\Pi_{VV'}$'s from the dark matter sector are given by,
\begin{align}
\Pi_{WW} =& -\frac{g^2}{16\pi^2} \sum_i \biggl( \left( |{\cal C}_{L,i}|^2 + |{\cal C}_{R,i}|^2  \right)H(m_i,m_D) + 4{\rm Re}({\cal C}_{L,i}{\cal C}_{R,i}^* ) m_D m_i B_0(m_i,m_D) \biggr), \\
\Pi_{ZZ} =& -\frac{g^2}{c^2}\frac{(1-2s^2)^2}{32\pi^2} \biggl( H(m_D,m_D) + 2 m_D^2 B_0(m_D,m_D) \biggr) \nonumber\\
& - \frac{g^2}{c^2}\frac{1}{32\pi^2} \sum_{i,j} \biggl( \left( |{\cal N}_{L,ij}|^2 + |{\cal N}_{R,ij}|^2  \right)H(m_i,m_j) + 4{\rm Re}({\cal N}_{L,ij}{\cal N}_{R,ji} ) m_i m_j B_0(m_i,m_j) \biggr), \\
\Pi_{Z\g} =& -\frac{eg}{c}\frac{1-2s^2}{16\pi^2}\biggl( H(m_D,m_D) + 2 m_D^2 B_0(m_D,m_D) \biggr), \\
\Pi_{\g\g} =& -\frac{e^2}{8\pi^2}\biggl( H(m_D,m_D) + 2 m_D^2 B_0(m_D,m_D) \biggr).
\end{align}
The definitions of $B_0$ and $H$ are given in the Appendix \ref{sec:oblique}.
By using the above two-point functions,
$S$ and $T$ are calculated by the formulae which are given in Eqs.~(\ref{eq:peskintakeuchi_s}, \ref{eq:peskintakeuchi_t}).
We show numerical results for $S$ and $T$ parameters in
Fig.~\ref{fig:st_f12}.

\subsubsection{Electric dipole moment}
\begin{figure}[p]
\centering
\includegraphics[width=5cm,angle=-90]{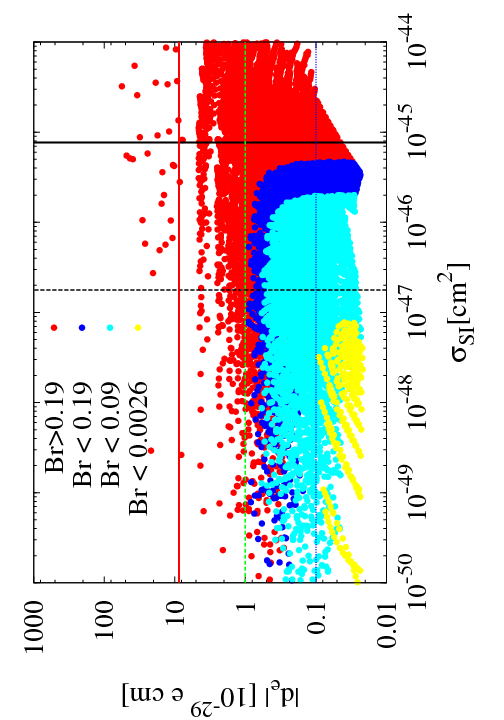}\\
\includegraphics[width=7cm]{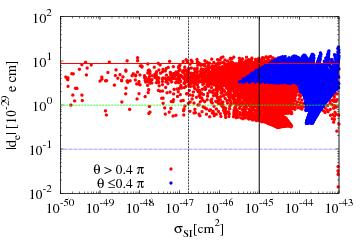}\\
\includegraphics[width=7cm]{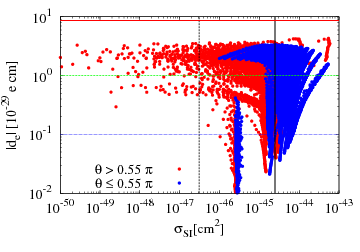}
\caption{
The spin independent cross section versus the electron EDM for $m_{\rm DM} = 40~\GEV$ (upper), $70~\GEV$ (middle) and $m_{\rm DM} = 200~\GEV$ (lower).
The red solid line is the current bound from ACME experiment.
The green and blue dashed lines are future
 prospects~\cite{Sakemi:2011zz,Kawall:2011zz}. 
The black solid line is the current bound by the LUX experiment. The black dashed line is the future prospect by XENON1T.  
}\label{fig:sigma_vs_edm}
\end{figure}

In model F12, as we have seen in Sec.~\ref{sec:model},
the Yukawa couplings of the dark matter can have a CP-violating phase,
and thus we
can probe dark matter sector by the measurement of EDMs.
In this model, two-loop diagram contributes to EDMs and its contribution is given by,\footnote{
We have checked that our calculation is consistent with Ref.~\cite{Giudice:2005rz}.}
\begin{align}
\frac{d_f}{e} = 2T_{3f} \sum_{i=1}^3 \left(\frac{g^2}{16\pi^2}\right)^2{\rm Im}[ {\cal C}_{L,i} {\cal C}_{R,i}^* ] \frac{m_i m_D m_f}{m_W^4}
\int_0^1 \frac{dx}{x} \frac{\log M_i^2(x)/m_W^2}{M_i^2(x)/m_W^2 - 1},
\end{align}
where $M_i^2(x) = m_i^2/(1-x) + m_D^2/x$. In the limit $m_D \gg m_S,~ yv, ~y'v$,
\begin{align}
\frac{d_f}{e} \simeq \frac{\a^2T_{3f}m_f}{64\pi^2 s_W^4} \frac{ {\rm Im}(yy')v^2 m_S}{m_W^2 m_D^3} \left( \log \frac{m_D^2}{m_W^2} + 1\right).
\end{align}
Numerically,
\begin{align}
d_e \simeq 8.6 \times 10^{-30} e~{\rm cm} \times {\rm Im}(yy')\left(\frac{m_S}{100~\GEV} \right)\left(\frac{m_D}{1000~\GEV} \right)^{-3} \left( \log\frac{m_D^2}{m_W^2} + 1\right).
\end{align}
Constraint on the electron EDM is given by the ACME experiment \cite{Baron:2013eja},
\begin{align}
|d_e| < 8.7\times 10^{-29} e~{\rm cm}\qquad(90\%~{\rm C.L.})
\end{align}
In Fig.~\ref{fig:mass-theta}, we show the non-zero CP violating phase opens large parameter space to avoid the constraints from the direct detection experiments.
The figure shows the electron EDM measurement is very useful to probe such a region.
We show the numerical result of electron EDM in Fig.~\ref{fig:sigma_vs_edm} with
future prospects~\cite{Sakemi:2011zz, Kara:2012ay,Kawall:2011zz}.

In the upper panel in Fig.~\ref{fig:sigma_vs_edm}, we also show the
branching fraction of the Higgs invisible decay. The red region is already
excluded at the LHC~\cite{Belanger:2013xza}. The blue region is within the
reach of the LHC~\cite{Baer:2013cma}.
The cyan region will be searched by the ILC experiment~\cite{Baer:2013cma}.
We find that the ILC has the capability to seek the parameter region where
both dark matter direct detection experiments and EDM experiments cannot access.

\subsubsection{Direct search}
\begin{figure}[t]
\includegraphics[width=0.48\hsize]{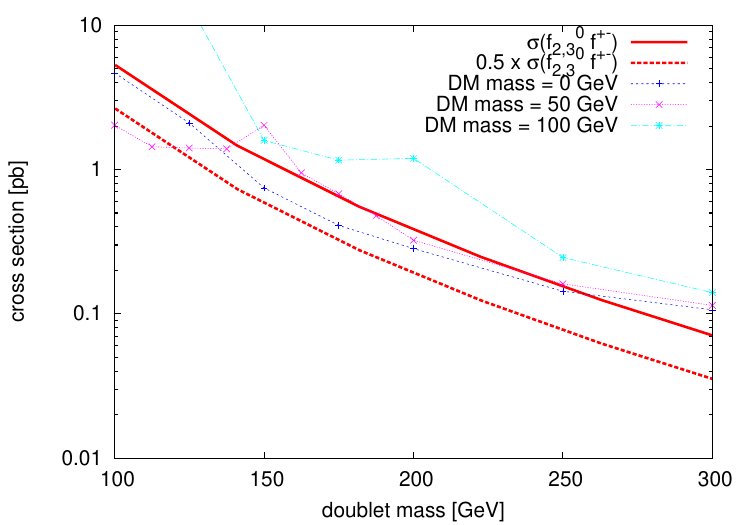}
\includegraphics[width=0.48\hsize]{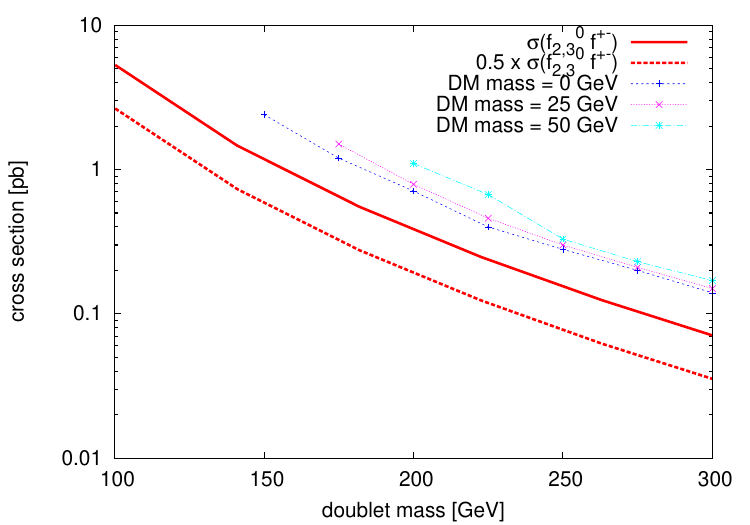}
\caption{
Constraint on production cross section of $f_{2,3}^0 f^{\pm}$ at the LHC $\sqrt s = 8~{\rm TeV}$.
Left figure shows the trilepton search \cite{ATLAS_Trilepton},
right figure shows the one lepton with two $b$-jets search \cite{ATLAS_OneleptonTwobjet}.
}\label{fig:f12_at_lhc}
\end{figure}

In the case $m_S, y v, y'v \ll m_D$, $f_2^0$, $f_3^0$ and $f^\pm$
approximately forms $SU(2)$ doublet, and their masses are almost degenerated.
The main decay modes of heavy matters are $f_{2,3}^0 \to f_1 (h/Z)$ and $f^\pm \to f_1 W^\pm$.
The main production channel at the LHC is $pp \to f_{2,3}^0 f^\pm \to f_1f_1 W^\pm (Z/h)$.
Such channels are searched in a context of electroweakino search in supersymmetric models.
The most sensitive channels are the trilepton mode \cite{ATLAS_Trilepton} and the one-lepton with two $b$-jets mode \cite{ATLAS_OneleptonTwobjet}.
The former mode makes the constraint on $\sum_i \s(pp\to f_i^0 f^\pm) {\rm Br}(f_i^0\to f_1Z ){\rm Br}(f^{\pm}\to f_1 W^{\pm} )$,
and the latter $\sum_i \s(pp\to f_i^0 f^\pm) {\rm Br}(f_i^0\to f_1h )
{\rm Br}(f^{\pm}\to f_1 W^{\pm} )$.
We estimated the production cross section by using Prospino2 \cite{Beenakker:1996ed} by taking pure Higgsino limit.
In Fig.~\ref{fig:f12_at_lhc}, we show the present status of constraint from direct search on model F12.
Ref.~\cite{ATL-PHYS-PUB-2014-010} shows that $m_{\rm wino} \lsim 800~{\rm GeV}$ can be probed
by trilepton search at LHC 14 TeV 3000 fb $^{-1}$ for wino, {\it i.e.}, $SU(2)$ triplet Majorana fermion.
Since the production cross section of $f_{2,3}^0 f^\pm$ in model F12 is a half of the cross section
for the pair production of a neutral wino and a charged wino,
we expect $m_D \lsim 600$--$700~{\rm GeV}$ can be probed LHC 14 TeV 3000 fb$^{-1}$.

\subsection{Model F23}\label{sec:F23}
Here, we discuss the phenomenology of model F23.
As we will see later, the measurement of the Higgs diphoton signal gives a severe constraint on the model. 

\subsubsection{Relic abundance and direct detection}
\begin{figure}[t]
\centering
\includegraphics[width=0.5\hsize]{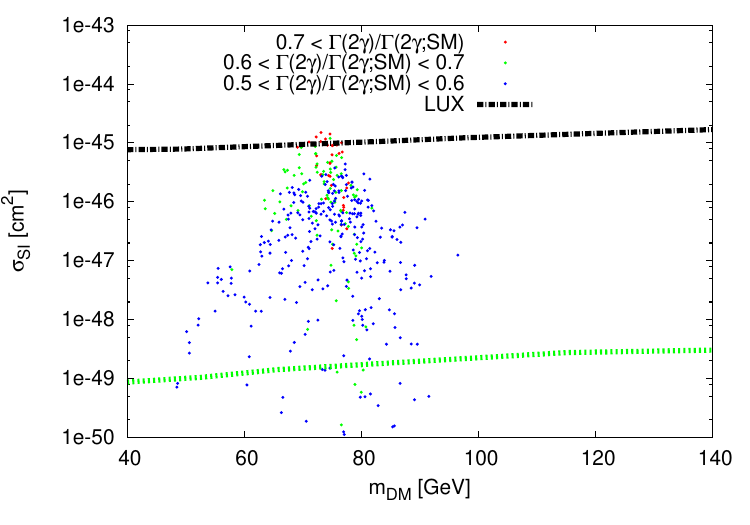}
\caption{
Scatter plot of $m_{\rm DM}$-$\s_{\rm SI}$ for model F23.
Black and green lines are same as Fig.~\ref{fig:mass-xsec-S2}.
Red points satisfies $\m(h\to 2\g) > 0.7$, and at green and blue points,
$0.6 < \m(h\to 2\g) < 0.7$, $0.5 < \m(h\to 2\g) < 0.6$, respectively.
}\label{fig:mass-xsec-f23}
\end{figure}

We show the dark matter mass and the spin independent cross section in Fig.~\ref{fig:mass-xsec-f23}.
In this plot, we take parameters of the model as
$\l, \l' \in [0,1.5]$, $ m_D, m_T \in [0,400]~\GEV$ and $\theta \in [0,\pi]$.
We calculate $\Omega_{\rm DM}h^2$, and extract the points which satisfy $0.1 \leq \Omega_{\rm DM} h^2 \leq 0.15$.

\subsubsection{Higgs diphoton decay signal}
\begin{figure}[t]
\centering
\includegraphics[width=0.48\hsize]{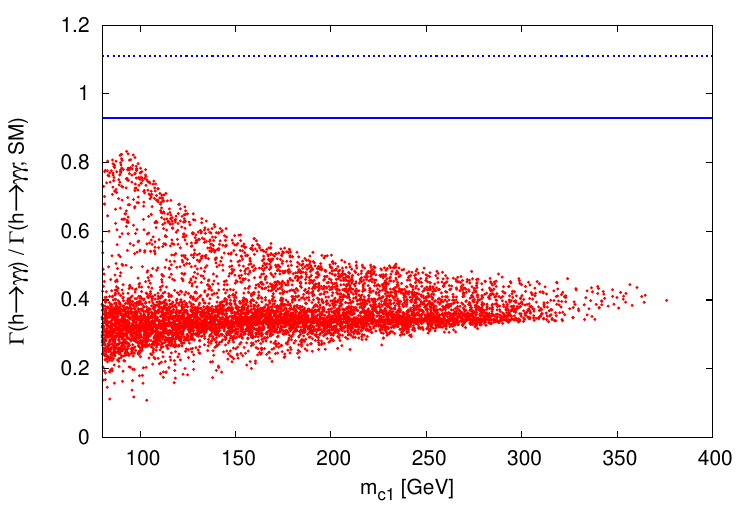}
\caption{
Diphoton signal strength and mass of the lightest charged fermion in model F23.
The region above dotted (solid) blue line is consistent with the measurement of $h\to\g\g$ within 1$\s$ (2$\s$) deviation.
}\label{fig:diphoton_f23}
\end{figure}

Here, we show the Higgs diphoton signal strength.
The interaction terms of charged fermion $\chi$'s are given by,
\begin{align}
{\cal L} = -y_{S,i} h \bar\Psi_i^+ \Psi_i -iy_{P,i} h \bar\Psi_i^+ \g^5 \Psi_i^+,
\end{align}
where the couplings $y_{S,i}$ and $y_{P,i}$ are determined by $\l$, $\l'$, $U^+$ and $U^-$ in Eq.~(\ref{eq:f23_mixing}).
The decay width of $h\to\g\g$ is given by,
\begin{align}
\G(h\to \g\g) =&
\frac{G_F \a^2 m_h^3}{128\sqrt 2 \pi^3} \left(
\left| A_{\rm SM}
+ \sum_i \frac{y_{S,i}v}{m_{\chi_i^\pm}} A_{1/2}^H\left( \frac{m_h^2}{4m_{\chi_i^\pm}^2} \right)\right|^2 
+ 
\left| \sum_i \frac{y_{P,i}v}{m_{\chi_i^\pm}} A_{1/2}^A \left( \frac{m_h^2}{4m_{\chi_i^\pm}^2} \right) \right|^2 \right).
\end{align}
We show how the diphoton branching fraction is modified in Fig.~\ref{fig:diphoton_f23}.
In this plot, we take the parameters of the model as
$\l, \l' \in [0,1.5]$, $ m_D, m_T \in [0,400]~\GEV$ and $\theta \in [0,\pi]$.
Then we calculate $\Omega_{\rm DM}h^2$ and extract the points which satisfy $0.1 \leq \Omega_{\rm DM} h^2 \leq 0.15$.
We use the constraint on $m_{\chi^+_1}$  by the chargino search at
the LEP experiment \cite{Heister:2002mn,Abbiendi:2002vz,Abdallah:2003gv,LEP2}, $m_{\chi^+_1} \lsim 93~\GEV$.
Applying this constraint, we can see from Fig.~\ref{fig:diphoton_f23}
that the diphoton signal strength is deviated from the SM value, $\m(h\to
2\g) < 0.85$.
However, the diphoton signal strength measured at the LHC is $\m(h\to 2\g) = 1.29 \pm 0.18$ (95~\% C.L.)\footnote{
This value is obtained from naive combination of
$1.65^{+0.33}_{-0.28}$ from the ATLAS collaboration \cite{ATLAS_diphoton}
and $1.14 \pm 0.21{\rm (stat.)} ^{+0.09}_{-0.05}{\rm (syst.)} ^{+0.13}_{-0.09}{\rm (theo.)}$ from the CMS collaboration \cite{Khachatryan:2014ira}. }. Therefore, this model is already excluded by LEP2 and LHC.
Hence, we do not investigate this model further in this paper.

\section{Discrimination of model S1 and F12}\label{sec:discriminate}
So far, we have discussed four dark matter models which are listed in Tab.~\ref{tab:modellist}.
Model S2 predicts $\sim$10~\% deviation of $\m(h\to 2\g)$ from the SM,
and the light mass region, $m_{\rm DM} < 72$~GeV, will be covered at the ILC.
On the other hand, model S1 and model F12 do not
predict a deviation in Higgs diphoton signal strength, and we can
distinguish them from model S2.
Model F23 predicts too large deviation of $\m(h\to 2\g)$ from the SM, and it
is already excluded.


The difference between model S1 and F12 is very subtle
because the phenomenology of dark matter in model S1 and F12 is quite similar.
If direct search experiments will discover the dark
matter, and if the dark matter mass and its spin independent cross
section are consistent with the prediction of model S1,
then we will have to discriminate model S1 from F12 by using some other
combination of observables. In this section, we discuss discrimination
of model S1 and F12 for each mass region.

\subsection{$m_{\rm DM} \lsim 53$~GeV}
In this mass region, 
as we can see from Fig.~\ref{fig:mass-xsec-fermion},
model S1 is already excluded by the dark matter direct search while model F12 is
not. Therefore we can distinguish these two models in this mass region by the
dark matter direct search.

\subsection{$53~\GEV \lsim m_{\rm DM} \lsim m_h / 2$}
\begin{figure}[p]
\centering
\includegraphics[width=0.48\hsize]{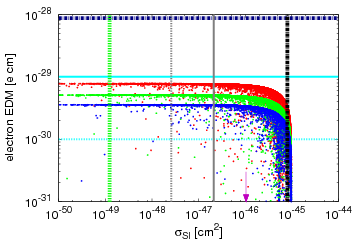}
\includegraphics[width=0.48\hsize]{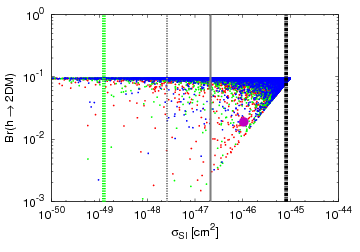}
\includegraphics[width=0.48\hsize]{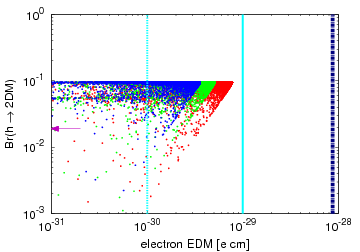}
\caption{
Discrimination of dark matter models for $m_{\rm DM} = 55~\GEV$.
Red, green, and blue points show $m_D = 200,~300$ and 400 GeV, respectively.
Magenta points or arrows show model S1.
Black chain line is constraint from the LUX, and gray solid and dotted line shows the future prospects of XENON1T and LZ experiment, respectively.
The values of experimental reach are taken from Ref.~\cite{Feng:2014uja}.
Green dotted line shows the discovery limit which is caused by atmospheric and astrophysical neutrinos.
Blue chain line shows the constraint from ACME experiment.
Solid turquoise line shows future prospect of measurement of Fr atom \cite{Sakemi:2011zz}.
Dotted turquoise line shows future prospect of measurement of YbF molecule and WN ion \cite{Kara:2012ay, Kawall:2011zz}.
}\label{fig:discrimination_55}
\end{figure}
\begin{figure}[t]
\centering
\includegraphics[width=0.48\hsize]{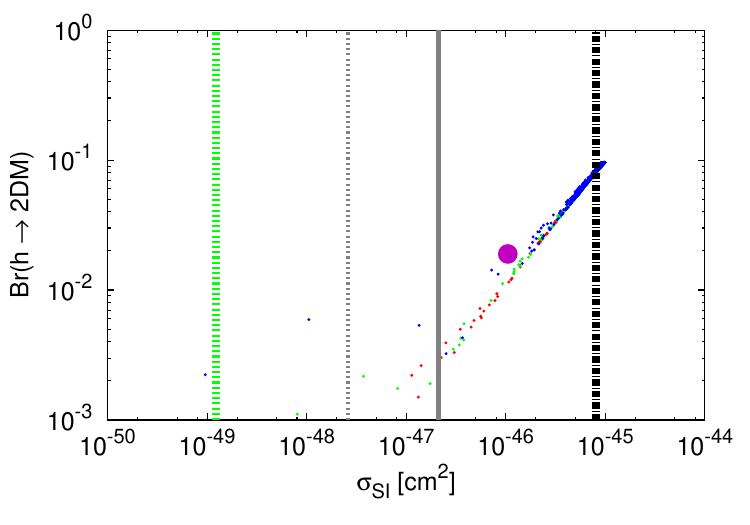}
\caption{Discrimination of dark matter models for $m_{\rm DM} = 55~\GEV$
in a parameter region with $|d_e| < 10^{-30} e{\rm cm}$, {\it i.e.},
in this case, future experiment \cite{Kara:2012ay, Kawall:2011zz} cannot
 observe electron EDM. 
The meaning of the lines and dot are the same as
 Fig.~\ref{fig:discrimination_55}. 
}\label{fig:discrimination_55_smallEDM}
\end{figure}
We show the correlations among the spin independent cross section, the electron EDM,
 and the branching fraction of the Higgs invisible decay in
 Fig.~\ref{fig:discrimination_55}.
Here we take $m_{\rm DM} = 55~\GEV$ as a benchmark. 
In these plots, for given $m_{\rm DM}$ and $m_D$, we take $|y/y'| \in
 [0,1]$ and $\theta \in [0,\pi]$ and take overall size of $y$ and $y'$
 to choose one which gives $\Omega_{\rm DM}h^2 = 0.1196$. In model F12
 with $m_D$ to be ${\cal O}(100)~\GEV$, this model gives various
 observables. Obviously, the electron EDM is a powerful tool for
 discrimination between S1 and F12 because model S1 does not include new
 CP violation source and does not predict any EDMs. 
We also show the case if future experiments \cite{Kara:2012ay, Kawall:2011zz}
do not observed electron EDM in Fig.~\ref{fig:discrimination_55_smallEDM}.
In this case, 
the branching fraction of the Higgs invisible decay is helpful to
 distinguish two models. 
The model F12 predicts wide range of the invisible width, while model S1
is a point.

We also check the case with $m_{\rm DM}=$60~GeV. In this case, both the electron
EDM and the branching fraction of the Higgs invisible decay are smaller than the
future prospect, and we have to rely on the direct search of the exotic
particles other than the dark matter particle in model F12 in order to
discriminate model S1 and F12.

\subsection{$100~\GEV \lsim m_{\rm DM}$}
\begin{figure}[t]
\centering
\includegraphics[width=0.48\hsize]{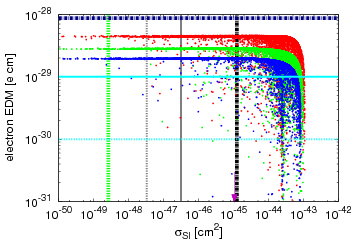}
\caption{
Scatter plot of $\s_{\rm SI}$--$d_e$ plane and $\s_{\rm SI}$--$S$ plane for $m_{\rm DM} = 100~\GEV$.
Red, green and blue points shows $m_D = 200,~300$ and $400~\GEV$, respectively.
Magenta points or arrows show model S1.
For the explanation of black, green, blue, turquoise lines, see the caption of Fig.~\ref{fig:discrimination_55}.
}\label{fig:discrimination_100}
\end{figure}
In the case of $m_{\rm DM} = 100~\GEV$, we show a scatter plot on the
$\s_{\rm SI}$--$d_e$ plane and $\s_{\rm SI}$--$S$ plane in
Fig.~\ref{fig:discrimination_100}. In these plots, for given $m_{\rm DM}$ and $m_D$, we take $|\l/\l'| \in [0,1]$ and $\theta \in [0,\pi]$
and take overall size of $\l$ and $\l'$ to choose one which gives $\Omega_{\rm DM}h^2 = 0.1196$.
Here, we can see that the electron EDM is very useful tools for the discrimination of the models.

\section{Conclusion and discussions}\label{sec:summary}
In this paper, we considered several simple dark matter models, and studied their phenomenological aspects comprehensively.
In particular, we discussed prospects of experimental reach to the dark matter models and discrimination of them
for the case of dark matter mass is smaller than ${\cal O}(100)~\GEV$.

\begin{figure}[t]
\centering
\includegraphics[width=\hsize]{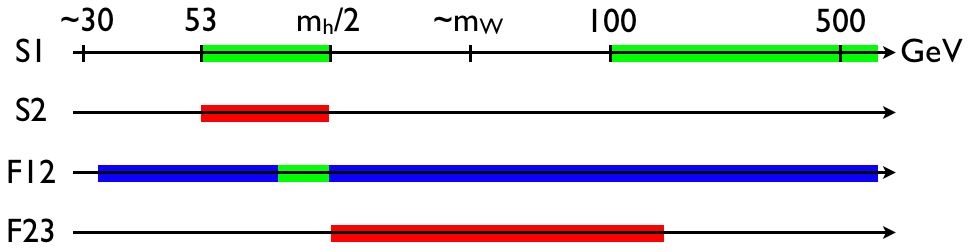}
\caption{Summary of the current status of each models in the light dark
 matter mass region. The color shaded
 regions are consistent with LUX experiments. The red shaded regions
 predict smaller diphoton signal strength. The blue shaded regions predict EDM.
}\label{fig:discrimination}
\end{figure}

\begin{figure}[t]
\centering
\includegraphics[width=0.70\hsize]{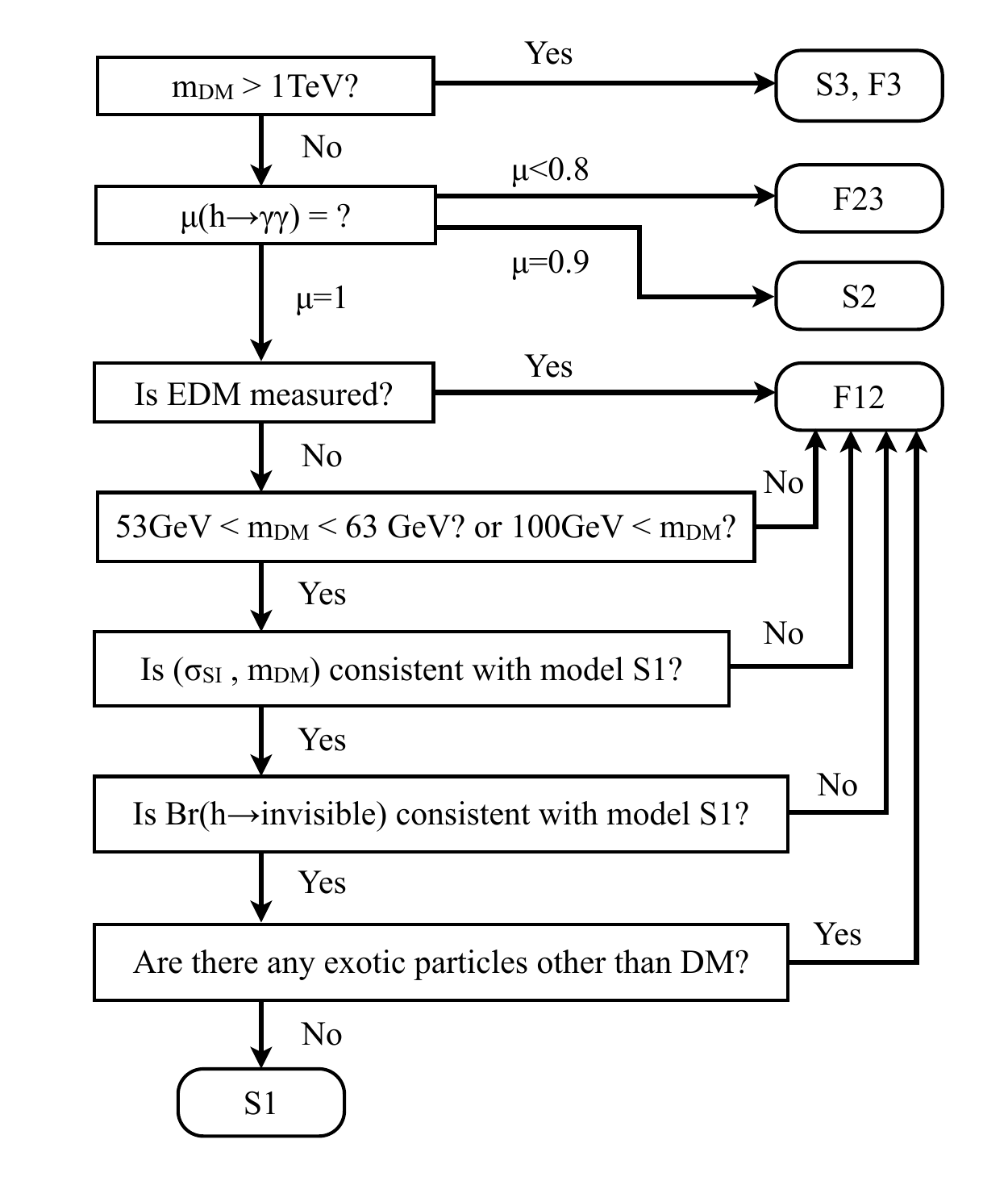}
\caption{Model chromatography for light dark matter models.
We consider model S1, S2, S3, F12, F23, and F3.
Here, we assume that only DM sector particles give contributions to electron EDM and deviation of diphoton signal strength from SM value.
}\label{fig:chromatgraphy}
\end{figure}

In this mass region, model S2 predicts 10\% deviation of $\m(h \to 2\g)$,
and thus the most of the region for the light dark matter in model S2
can be covered by the LHC and the ILC.
Model F23 predicts large $\m(h\to 2\g)$ deviation and already excluded.
For model F12, in the case of the doublet Dirac mass $m_D$ to be a few hundred GeV, 
the observation of the electron EDM and the discovery of the direct search for doublet fermions are expected.
For $ 53~\GEV < m_{\rm DM} < m_h / 2$ and $100~\GEV < m_{\rm DM}$, 
if the electron EDM is not observed, it is not easy task to distinguish
model S1 and F12,
because the spin-independent cross section for model F12 can mimic the
one for model S1 due to the existence of the blind spots. The
measurements of the branching fraction of the Higgs invisible decay at the ILC provide
us with useful information in the case. Of course, the direct
search for other $Z_2$ odd particles is also useful to distinguish model
S1 and F12.
We summarize the features of each models for light dark matter in 
Tab.~\ref{tab:last_table}, and current status of the dark matter mass region in 
Fig.~\ref{fig:discrimination}.
We also summarize how to distinguish light dark matter models which we
addressed in this paper in Fig.~\ref{fig:chromatgraphy}.
\begin{table}
\centering
\caption{ Summary of light dark matter. The cells marked ``-'' are not
 treated in this paper.  }
\begin{tabular}{|c||c|c|c|c|c|}
\hline
 & S1 & S2 & F12 & F23 \\
\hline
\hline
$\m(h \to 2\g)$        & 1 (same as SM) & $\sim 0.9$     & 1 (same as SM)                  & $\lsim 0.8$ \\ \hline
EWPT                   & (same as SM)   &                &                                 & - \\ \hline
EDM                    & (same as SM)   & (same as SM)   & $> 10^{-30} e$
	     cm is possible      & - \\ \hline
Collider               & -              & -              & LHC                             & - \\ \hline
\end{tabular}
\label{tab:last_table}
\end{table}

\section*{Acknowledgements}
This work is supported by JSPS Grant-in-Aid for Young
Scientists (B) (No.~23740165 [RK]), MEXT Grant-in-Aid
for Scientific Research on Innovative Areas (No.~25105011 [RK] 
and No.~23104006 [TA]) and JSPS
Research Fellowships for Young Scientists [RS].

\appendix
\section{Loop functions}
Here, we summarize the loop functions which are used in the calculation of EWPT and diphoton signal strength.

\subsection{Loop functions for oblique corrections} \label{sec:oblique}
Scalar loop contribution to vector boson two-point function is given as,
\begin{align}
& \m^{4-d} \left[ \int \frac{d^d k}{(2\pi)^d} \frac{ (2k+p)^\m (2k+p)^\n }{ [(k+p)^2-m_1^2] [k^2-m_2^2] }
 + \int \frac{d^d k}{(2\pi)^d} \frac{g^{\m\n}}{k^2-m_1^2}
 + \int \frac{d^d k}{(2\pi)^d} \frac{g^{\m\n}}{k^2-m_2^2} \right] \nonumber\\
=& 
\frac{i}{16\pi^2} 4 \tilde B_{22} (m_1,m_2) g^{\m\n} 
+ 
(p^{\mu} p^{\nu} \textrm{ terms})
.
\label{eq:sloop}
\end{align}
Fermion loop contribution to vector boson two-point function is given as,
\begin{align}
& (-1) \m^{4-d} \int \frac{d^d k}{(2\pi)^d} \int {\rm tr}\left[
 \g^\m(g_L P_L + g_R P_R) \frac{ \sla{k}+\sla{p}+m_1 }{ (k+p)^2-m_1^2 }
 \g^\m(h_L P_L + h_R P_R) \frac{ \sla{k}+m_2 }{ k^2-m_2^2 } \right] \label{eq:floop} \nonumber\\
=& \frac{-i}{16\pi^2}\biggl[ (g_L h_L + g_R h_R) H(m_1,m_2) + 2(g_L h_R + g_R h_L) m_1 m_2 B_0(m_1,m_2) \biggr]g^{\m\n} + \cdots.
\end{align}
Here, $\cdots$ represents terms which are proportional to $p^\m p^\n$.
If internal fermion is same Majorana fermion, Eq. (\ref{eq:floop})
accompanies with symmetric factor $1/2$. 
$B_0$, $H$ and $\tilde B_{22}$ are loop functions given in
Ref. \cite{Pierce:1996zz}, and their integral forms are, 
\begin{align}
B_0(m_1,m_2) &= \frac{1}{\hat\e} - \int_0^1 dx \log \frac{\D}{\m^2}, \\
H(m_1,m_2) &= \frac{1}{\hat\e}\left(\frac{2}{3}p^2 - m_1^2 - m_2^2 \right) - \int_0^1 dx \biggl( 4x(1-x)p^2 - 2xm_1^2 - 2(1-x)m_2^2  \biggr) \log \frac{\D}{\m^2}, \\
\tilde B_{22}(m_1,m_2) &= -\frac{1}{\hat\e}\frac{p^2}{12} + \frac{1}{4}\int_0^1 dx \biggl( (1-2x)^2 p^2 - (m_1^2-m_2^2)(1-2x) \biggr) \log \frac{\D}{\m^2}.
\end{align}
where $\D = x m_1^2 + (1-x) m_2^2 - x(1-x)p^2 - i\e$,
 $\hat\e = 2-d/2$ 
and 
$\m$ 
is the renormalization scale in $\overline{\rm MS}$ scheme.

\subsection{Loop functions for diphoton signal}\label{sec:diphotonloopfunc}
$A$'s are loop functions which are defined in Ref. \cite{Djouadi:2005gi}.
They are defined as,
\begin{align}
A_1(\t) &= -\t^{-2}( 2\t^2+3\t+3(2\t-1)f(\t) ), \\
A_{1/2}^A(\t) &= 2\t^{-1}f(\t),\\
A_{1/2}^H(\t) &= 2\t^{-2}(\t + (\t-1)f(\t)),\\
A_0(\t) &= -\t^{-2}( \t-f(\t) ), 
\end{align}
where $f(\t)$ is defined as,
\begin{align}
f(\t) = \left\{
\begin{array}{ll}
\arcsin^2\sqrt{\t}. & (\t \leq 1) \\
-\displaystyle\frac{1}{4}\left(\log\displaystyle\frac{1+\sqrt{1-\t^{-1}}}{1-\sqrt{1-\t^{-1}}} - i\pi \right)^2. & (\t > 1)
\end{array}
\right.
\end{align}


\begin{thebibliography}{99}
\bibitem{CMS:2014ega} 
  CMS Collaboration [CMS Collaboration],
and studies of the compatibility of its couplings with the standard model,''
  CMS-PAS-HIG-14-009.
\end{thebibliography}
\end{document}